\PassOptionsToPackage{dvipsnames}{xcolor}
\documentclass[longbibliography,nofootinbib,aps,prx,superscriptaddress,twocolumn]{revtex4-2}

\usepackage{amsmath,amsfonts,amssymb,amsthm,dcolumn,bm,qcircuit,appendix,mathtools,thmtools,thm-restate,algorithm,algpseudocode,mathrsfs,pgfplots,soul,lipsum,graphicx,braket,enumitem,caption,subcaption,ragged2e}

\usepackage[
  labelfont=bf,
  labelsep=period,
  justification=justified,
  format=plain,
  singlelinecheck=false
]{caption}

\usepackage{pgfplots}
\usepgfplotslibrary{patchplots}
\usetikzlibrary{3d, decorations.pathmorphing}
\usepackage{tikz,tkz-euclide,tikz-3dplot}
\pgfplotsset{compat=newest}
\usepgfplotslibrary{colormaps}
\usepgfplotslibrary{patchplots}

\usetikzlibrary{matrix,fit,backgrounds,3d,arrows, decorations.pathreplacing,shapes.geometric,3d, calc}

\def\BibTeX{{\rm B\kern-.05em{\sc i\kern-.025em b}\kern-.08em
    T\kern-.1667em\lower.7ex\hbox{E}\kern-.125emX}}

\newtheorem{proposition}{Proposition}
\newtheorem{theorem}{Theorem}
\newtheorem{lemma}{Lemma}

\newtheorem{remark}{Remark}
\newtheorem{definition}{Definition}
\newtheorem{method}{Algorithm}

\numberwithin{proposition}{section}
\numberwithin{theorem}{section}
\numberwithin{lemma}{section}
\numberwithin{corollary}{section}
\numberwithin{remark}{section}
\numberwithin{definition}{section}
\numberwithin{equation}{section}

\DeclareMathOperator{\Tr}{Tr}
\newcommand{\xbf}{\textbf{x}}

\newcommand{\Ibb}{\mathbb{I}}

\newcommand{\Rbb}{\mathbb{R}}

\newcommand{\norm}[1]{\left\lVert#1\right\rVert}

\usepackage{xcolor}
\usepackage{hyperref}

\hypersetup{
    colorlinks=true,
    linkcolor=MidnightBlue,
    filecolor=magenta,
    urlcolor=teal,
    pdfpagemode=FullScreen,
    citecolor=OliveGreen
}

\begin{document}
\title{Quantum Algorithm for Estimating Intrinsic Geometry}

\author{Nhat A. Nghiem}
\email{{nhatanh.nghiemvu@stonybrook.edu}}
\affiliation{QuEra Computing Inc., Boston, Massachusetts 02135, USA}
\affiliation{Department of Physics and Astronomy, State University of New York at Stony Brook, \\ Stony Brook, NY 11794-3800, USA}
\affiliation{C. N. Yang Institute for Theoretical Physics, State University of New York at Stony Brook, \\ Stony Brook, NY 11794-3840, USA}

\author{Tuan K. Do}
\email{{ktdo@ucsb.edu}}
\affiliation{Department of Mathematics,  University of California, Santa Barbara, CA
93106, USA}

\author{Tzu-Chieh Wei}
\email{{tzu-chieh.wei@stonybrook.edu}}
\affiliation{Department of Physics and Astronomy, State University of New York at Stony Brook, \\ Stony Brook, NY 11794-3800, USA}
\affiliation{C. N. Yang Institute for Theoretical Physics, State University of New York at Stony Brook, \\ Stony Brook, NY 11794-3840, USA}

\author{Trung V. Phan}
\email{{tphan@natsci.claremont.edu}}
\affiliation{Department of Natural Sciences, Scripps and Pitzer Colleges, \\ Claremont Colleges Consortium, Claremont, CA 91711, USA}

\begin{abstract}
High-dimensional datasets typically cluster around lower-dimensional manifolds but are also often marred by severe noise, obscuring the intrinsic geometry essential for downstream learning tasks. We present a quantum algorithm for estimating the intrinsic geometry of a point cloud -- specifically its local intrinsic dimension and local scalar curvature. These quantities are crucial for dimensionality reduction, feature extraction, and anomaly detection -- tasks that are central to a wide range of data-driven and data-assisted applications. In this work, we propose a quantum algorithm which takes a dataset with pairwise geometric distance, output the estimation of local dimension and curvature at a given point. We demonstrate that this quantum algorithm achieves an exponential speedup over its classical counterpart, and, as a corollary, further extend our main technique to diffusion maps, yielding exponential improvements even over existing quantum algorithms. Our work marks another step toward efficient quantum applications in geometrical data analysis, moving beyond topological summaries toward precise geometric inference and opening a novel, scalable path to quantum-enhanced manifold learning.
\end{abstract}

\maketitle

\section{Introduction}

\begin{figure*}[tb!]
    \centering
    \includegraphics[width=0.8\linewidth]{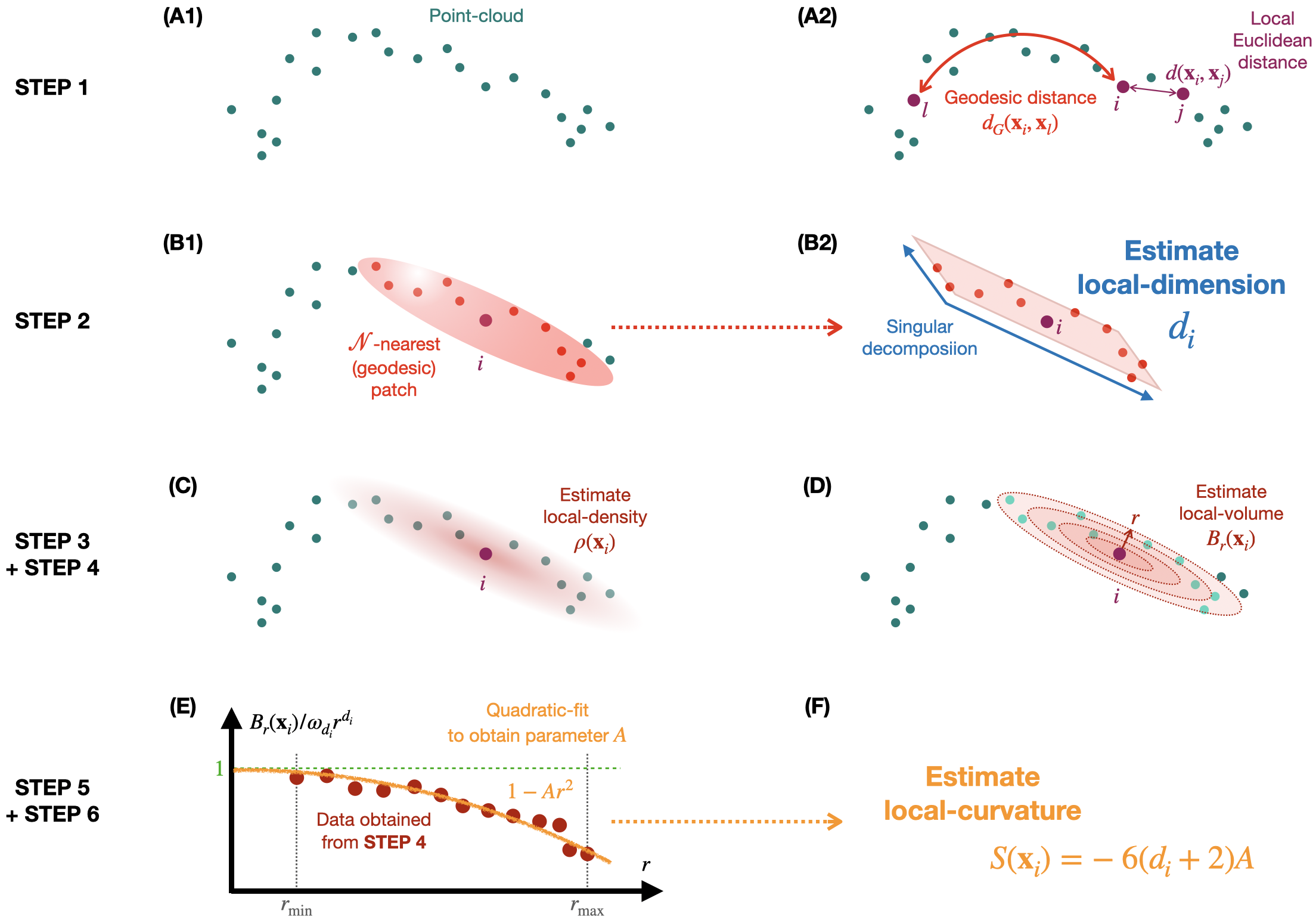}
    \caption{\textbf{An illustration of the classical algorithm for estimating intrinsic dimension and scalar curvature.} The details of this algorithm can be found in Section \ref{sec: classicalalgorithm} and Appendix \ref{sec: pipelineclassicalalgorithm}. \textbf{(A)} \underline{Step 1:} Starting with a noisy point cloud \textbf{(A1)}, we use diffusion geometry to define geodesic distances between any pair of points \textbf{(A2)}. \textbf{(B)} \underline{Step 2:} We select a local neighborhood of nearest points geodesically \textbf{(B1)} to perform PCA and estimate the intrinsic dimensionality \textbf{(B2)}. \textbf{(C)} \underline{Step 3:} We estimate the sampling density on the manifold, using a heat-kernel smoothing. \textbf{(D)} \underline{Step 4:} We estimate the volume of a geodesic ball via importance sampling for different radii. \textbf{(E)} \underline{Step 5:} We do a function fit to the data obtained from \underline{Step 4}. \textbf{(F)} \underline{Step 6:} We estimate the local scalar-curvature from the best-fit parameter found in \underline{Step 5}.}
    \label{fig:curvatureestimation}
\end{figure*}
 
It is commonly assumed under the \textit{manifold hypothesis}~\cite{fefferman2016testing,gorban2018blessing,gorban2018blessing,brown2022verifying} that most real-world high-dimensional data reside in a lower-dimensional manifold. Dimensional reduction removes redundant embedding---often due to noise or extrinsic constraints~\cite{sternad2018s}---and reveals the \textit{true} degrees of freedom that capture the underlying geometry and variability of the system \textit{locally}. The number of these local degrees is known as the \textit{intrinsic dimension}~\cite{amsaleg2015estimating}, and accurately estimating it helps to denoise the data efficiently, reduce storage and computational costs, while still preserving meaningful geometric structures. Importantly, estimating the intrinsic dimension \textit{locally} provides a direct way to assess whether the manifold hypothesis holds. For machine learning applications, the existence of a well-defined \textit{global} intrinsic dimension informs the \textit{embedding dimension}---the minimum number of input features needed to describe the system manifold \textit{globally}---upper‑bounded by Whitney's embedding theorem~\cite{whitney1936differentiable,lee2003smooth}. Once the global intrinsic dimension is known, one can then estimate the \textit{scalar-curvature}~\cite{lee2006riemannian}, which has been rigorously proven to be a noise- and sampling-robust measure that quantifies the local shape of the manifold~\cite{hickok2023intrinsic}. Curvature-based diagnostics enable uncovering anomalies, such as bottlenecks or singularities, and in doing so provide actionable feedback for further data collection by flagging regions where additional or higher-resolution sampling would be most informative~\cite{grover2025curvgad}. Thus, estimating the intrinsic dimension and the scalar curvature can bridge between raw observations and geometry-aware analysis, as these quantities are essential in both \textit{data-driven} pipelines, where models are learned directly from data, and \textit{data-assisted} workflows, where data refine or constrain physics-based models.

While classical algorithms for manifold learning are well-developed and widely used~\cite{jones2024manifold}, their quantum counterparts remain largely unexplored. Recent progress has shown promise for quantum computers in tackling problems in \textit{topological data analysis} (TDA), where tools such as persistent homology, Betti number estimation, and homology class tracking are employed to extract robust \textit{topological invariants} from data~\cite{lloyd2016quantum,ubaru2021quantum,schmidhuber2022complexity,nghiem2023quantum, berry2024analyzing, lee2025new}. Motivated by these advances, we turn our attention to the potential of quantum computing for \textit{geometric data analysis} (GDA). In contrast to TDA, which focuses on global topological features, GDA aims to uncover the underlying geometric structure of data (e.g., distances, angles, and curvatures), offering a more localized and fine-grained description that is especially important in modeling data lying near continuous or manifold-structured domains.

Quantum approaches to GDA are still in their infancy. One of the key challenges lies in estimating the \textit{geodesic distances} from raw pairwise separations---a crucial step in recovering the intrinsic manifold geometry, where geodesic paths are often induced via \textit{diffusion geometry} by constructing a diffusion operator on the data point cloud that converges to the Laplace–Beltrami operator on the underlying manifold~\cite{coifman2006diffusion,belkin2003laplacian,jones2024manifold}. Classically, this can be approximated using kernel methods, with geometric information encoded in the spectrum and eigenvectors of the kernel matrix. However, extracting the geodesic structure from quantum-accessible representations of such kernels---particularly from their eigenvectors---is not trivial. In this work, we propose an approach that integrates the kernel approximation with a quantum routine designed to recover relevant quantities more directly and \textit{exponentially faster} than its classical counterpart. Here, as a demonstration, we apply it to build a quantum algorithm that estimates intrinsic dimension and scalar curvature with an exponential speed-up over its classical counterpart. More generally, our approach provides a novel pathway toward quantum estimation of a broader class of intrinsic geometric quantities, laying the foundation for future developments in quantum GDA. In particular, applying our technique to diffusion maps also yields an \textit{exponential improvement} over a previously proposed quantum algorithm~\cite{sornsaeng2021quantum}.

Our work is organized as follows. Section \ref{sec: classicalalgorithm} reviews the classical estimator for local intrinsic dimension and curvature, following \cite{hickok2023intrinsic}. Section \ref{sec: quantumalgorithm} presents our quantum algorithm, summarized in Fig.~\ref{fig: Diagram}, with technical details deferred to the Appendix. Section \ref{sec: discussion} contextualizes our findings, highlighting the quantum advantage and an extension to diffusion maps.

\section{Classical Algorithm}
\label{sec: classicalalgorithm}

In this Section, we review the classical algorithm for estimating the intrinsic dimension via singular value decomposition (SVD) and the scalar curvature through a noise-resistant geometric formulation proposed in~\cite{hickok2023intrinsic}. We begin with the theoretical foundations on a smooth differentiable Riemannian manifold $(\mathcal{M},g)$, where $\mathcal{M}$ is a $d$-dimensional manifold and $g$ is a symmetric positive-definite Riemannian metric tensor that allows us to define geometric notions, e.g., lengths of curves, angles between vectors, and volumes of subsets on the manifold. A ball $B_r(p)$ is defined as the collection of points whose geodesic distance from the point $p \in \mathcal{M}$ is less than the length $r$, which we refer to as the geodesic radius of the ball. The volume of $B_r(p)$ satisfies the following expansion in the powers of the radius $r$ \cite{gray1979riemannian}:
\begin{equation}
    \frac{\text{Vol}[B_r(p)]}{\omega_d r^d} \approx 1 - \frac{S(p)}{6(d+2)} r^2 + \mathcal{O}(r^4) \ ,
\label{big_ball}
\end{equation}
where $\omega_d = \pi^{d/2}/\Gamma(d/2+1)$ with $\Gamma(x)$ being the gamma function and $S(p)$ is the scalar curvature evaluated at point $p$. The factor $\omega_d r^d$ is the volume of a $d$-dimensional Euclidean ball of radius $r$; if the manifold were perfectly flat near $p$, the volume ratio in Eq. \eqref{big_ball} would be $1$. The curvature ``bends geodesics'' either toward each other (positive scalar curvature) or away from each other (negative scalar curvature), so the actual geodesic ball becomes slightly smaller or larger, respectively, than its Euclidean counterpart. The leading-order correction is proportional to the scalar curvature $S(p)$; a positive $S(p)$ decreases the volume, whereas a negative $S(p)$ increases it. Higher-order terms of order $r^4$ and beyond capture finer geometric effects but vanish rapidly as the radius shrinks. We give a derivation for this formula in Appendix~\ref{sec:ball_formula} and refer the readers who are unfamiliar with differential geometry to Appendix~\ref{sec: reviewofdifferentialgeomtry} for an overview of the subject.

The classical algorithm for estimating the scalar curvature at a point within a point cloud is based on Eq.~\eqref{big_ball}~\cite{gray1979riemannian}. However, before applying this formula, one must first define the geodesic distance and estimate the intrinsic dimension of the manifold that effectively approximates the noisy point cloud (see Fig.~\ref{fig:curvatureestimation}A1):
\begin{itemize}
    \item To define the geodesic distance, we require a notion of \textit{continuity}. This can be established by diffusion geometry, in which Euclidean separations between points are converted into smooth affinity weights. By seeding a diffusion process with those affinities and measuring how influence propagates across the point-cloud, one can then approximate the underlying manifold's true geodesic distances~\cite{coifman2006diffusion} (see Fig.~\ref{fig:curvatureestimation}A2).
    \item To estimate the local intrinsic dimension $d_i$ around a point ${\bf x}_i$ in the point-cloud, we can define \textit{a neighborhood} as the set of $\mathcal{N}$-closest points geodesically to ${\bf x}_i$, including itself (see Fig.~\ref{fig:curvatureestimation}B1). The value $\mathcal{N}$ serves as a \textit{hyperparameter} that controls the locality of the estimate. Assume the manifold is locally-flat, \textit{principal component analysis} (PCA)---a linear method based on SVD---can then be applied to this neighborhood to recover the intrinsic dimensionality $d_i$ (see Fig.~\ref{fig:curvatureestimation}B2). The \textit{global} intrinsic dimension $d$\footnote{While it is worth noting that the \textit{global} intrinsic dimension can also be estimated directly by analyzing how affinity information spreads across the entire point cloud~\cite{coifman2006diffusion}, we consider the local approach here, as it provides a richer, spatially resolved description that can capture heterogeneity and validate of the manifold hypothesis more thoroughly.}---representing the dimensionality of the point cloud underlying manifold---is most likely given by the median of all \textit{local} estimates $d_i$.
\end{itemize}
If the total number of points in the point cloud is $N$, a good choice for $\mathcal{N}$ should obeys the numerical scale hierarchy $d < \mathcal{N} \ll N$, ensuring that every local connection spans the $d$-dimensional manifold tangent space while each patch is large enough for stable PCA yet still small enough to stay within the locally-flat regime.

If the data were drawn from a uniform distribution on $\mathcal{M}$, simply counting the number of points inside a geodesic ball $B_r({\bf x}_i)$ would give an unbiased proxy for its volume. However, real data are almost always sampled with an unknown, spatially varying density $\rho({\bf x})$, thus mixing geometric volume with the unknown sampling density, which makes it a poor estimator. To disentangle these two effects:
\begin{itemize}
    \item We need to estimate the sampling density $\rho({\bf x})$, which can be done using a heat-kernel smoothing method (see Fig.~\ref{fig:curvatureestimation}C).
    \item Then, we can perform an \textit{importance sampling} by weighting each point inversely by the estimated density, so that sparsely-sampled regions contribute more and densely-sampled regions contribute less. To see how this weighted-counting recovers an unbiased estimator of the geometric volume, we note that:
    \begin{equation}
    \begin{split}
        \mathbb{E}\left[\sum_{{\bf x}_j \in B_r ({\bf x}_i) } \rho^{-1}({\bf x}_j) \right] &= \int_{B_r ({\bf x}_i)} \rho({\bf x}_i) dV \rho^{-1}({\bf x}_i) 
        \\
         = \int_{B_r ({\bf x}_i)} dV &= \text{Vol}[B_r ({\bf x}_i)]  \ ,
    \label{estimate_volume}
    \end{split}
    \end{equation}
    where $\mathbb{E}(\circ)$ denotes the statistical-expectation of the quantity $\circ$. This estimator can be used to measure $\text{Vol}[B_r ({\bf x}_i)]$ across varying geodesic radii $r$ (see Fig.~\ref{fig:curvatureestimation}D).
\end{itemize}
Note that we have introduce two more \textit{hyperparameters}, $r_{\min}$ and $r_{\max}$, to investigate the $r$-dependence of $\text{Vol}[B_r ({\bf x}_i)]/\omega_d r^d$ and fit it with an one degree of freedom function $1+Ar^2$ (see Fig.~\ref{fig:curvatureestimation}E). A good choice for the lower-threshold $r_{\min}$ should be about the typical geodesic distance between points in a neighborhood, ensuring the analysis remains above the noise level while still capturing fine geometric features. The upper-threshold $r_{\max}$ should be set only a few times higher than $r_{\min}$, so that higher-order corrections in Eq.~\eqref{big_ball} stay small. The best-fit value of $A$, obtained by minimizing the squared deviation of the fit across the interval $r \in [r_{\min},r_{\max}]$, gives us a noise-resistant estimation for the local scalar curvature $S({\bf x}_i)$ \cite{hickok2023intrinsic} (see Fig.~\ref{fig:curvatureestimation}F).

We present a more detailed explanation for the classical algorithm than described here in the Appendix~\ref{sec: pipelineclassicalalgorithm}. 

\section{Quantum Algorithm}
\label{sec: quantumalgorithm}

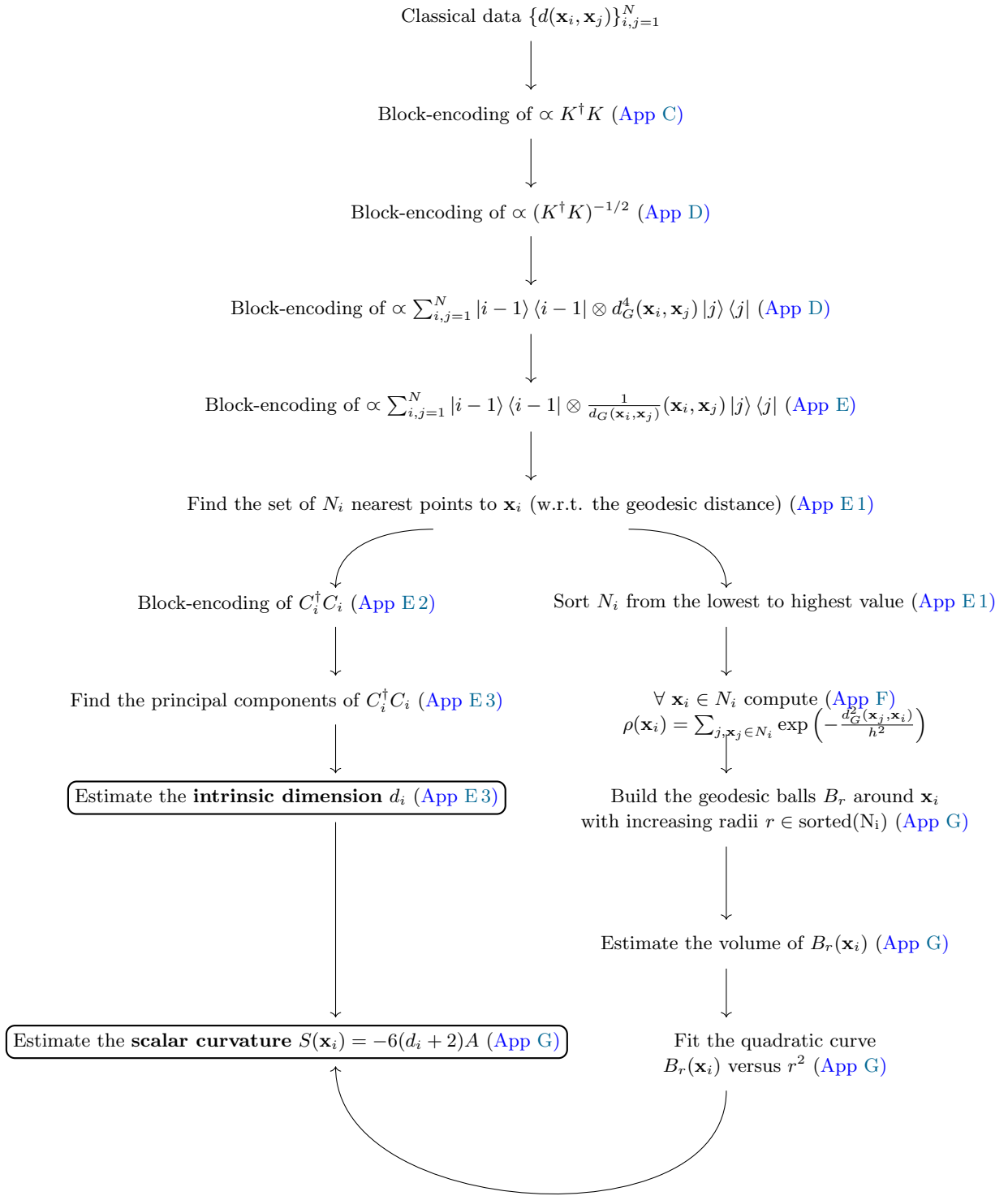
\begin{figure*}[t!]
   \centering
\begin{tikzpicture}[scale = 0.8]
    \node at (0,5) {Classical data $\{ d(\xbf_i,\xbf_j)\}_{i,j=1}^N$} ;
 \draw[->] (0,4.5) -- (0,3.5) ; 
    \node at (0,3) {Block-encoding of $\varpropto K^\dagger K$ (\color{blue}{App \ref{sec: blockencodingkernelmatrix})}};
 \draw[->] (0,2.5) -- (0,1.5) ;
    \node at (0,1) {Block-encoding of $\varpropto (K^\dagger K)^{-1/2}$ (\color{blue}{App \ref{sec: detailgeodesicdistance})}   } ;
 \draw[->] (0,0.5) -- (0,-0.5) ;
    \node at (0,-1) {Block-encoding of $\varpropto \sum_{i,j=1}^N\ket{i-1}\bra{i-1} \otimes d_G^4(\xbf_i,\xbf_j) \ket{j}\bra{j}  $ (\color{blue}{App \ref{sec: detailgeodesicdistance})}}; 
 \draw[->] (0,-1.5) -- (0,-2.5) ;
    \node at (0,-3) {Block-encoding of $\varpropto \sum_{i,j=1}^N\ket{i-1}\bra{i-1} \otimes \frac{1}{d_G(\xbf_i,\xbf_j)}(\xbf_i,\xbf_j) \ket{j}\bra{j}  $ (\color{blue}{App \ref{sec: detailestimatingintrinsicdistance})}};
     \draw[->] (0,-3.5) -- (0,-4.5) ;
    \node at (0,-5) { Find the set of $N_i$ nearest points to $\xbf_i$ (w.r.t. the geodesic distance) (\color{blue}{App \ref{sec: findingninearestpoints})} } ; 
    \draw[->] (-2, -5.5)  to[out = 180, in = 90 ] (-4, -6.7);
\draw[->] (2, -5.5)  to[out = 0, in = 90 ] (4, -6.7);
    \node at (-5, -7) {Block-encoding of $C_i^\dagger C_i$ (\color{blue}{App \ref{sec: obtainblockencodingofcici})}};
    \draw[->] (-4, -7.5) -- (-4, -8.5);
    \node at (5, -7) { Sort $N_i$ from the lowest to highest value (\color{blue}{App \ref{sec: findingninearestpoints})} };
     \draw[->] (4, -7.5) -- (4, -8.5);
    \node at (-5, -9) {Find the principal components of $C_i^\dagger C_i$ (\color{blue}{App \ref{sec: estimatinglocalintrinsicdimension})}};
    \node at (5, -9) { $\forall \ \xbf_i \in N_i$ compute (\color{blue}{App \ref{sec: quantumalgorithmestimatingdensity})}};
    \node at (5, -9.5) { $\rho(\xbf_i) = \sum_{j, \xbf_j \in N_i} \exp\left(- \frac{d_G^2(\xbf_j,\xbf_i)}{h^2} \right) $};
    \node[draw, thick, rounded corners] at (-5, -11) { Estimate the \textbf{intrinsic dimension} $d_i$ (\color{blue}{App \ref{sec: estimatinglocalintrinsicdimension})}};
    \node at (5, -11) { Build the geodesic balls $B_r$ around $\xbf_i$}; 
    \node at (5, -11.5) {with increasing radii $r \in \rm sorted( N_i)$ (\color{blue}{App \ref{sec: fitcurvature})} };
    \draw[->] (4, -12) --  (4, -13.5);
    \node at (5, -14) {Estimate the volume of $B_r(\xbf_i)$ (\color{blue}{App \ref{sec: fitcurvature})}  };
    \draw[->] (-4, -9.5) -- (-4, -10.5);
    \draw[->] (4, -9.7) -- (4, -10.5); 
    \draw[->] (4, -14.5) -- (4, -15.5) ;
    \node at (5, -16)  {Fit the quadratic curve }; 
    \node at (5, -16.5) { $B_r(\xbf_i)$ versus $r^2$ (\color{blue}{App \ref{sec: fitcurvature})} }; 
    \node[draw, thick, rounded corners] at (-5, -16) {Estimate the \textbf{scalar curvature} $S(\xbf_i) = -6 (d_i+2) A$ (\color{blue}{App \ref{sec: fitcurvature})}};
    \draw[->] (-4, -11.5) -- (-4, -15.5);
    \draw[->] (4, -17)   to[out = 270, in = 270] (-4, -16.5) ;
\end{tikzpicture}
    \caption{The work-flow of our quantum algorithm for estimating the intrinsic dimension $d$ and local scalar curvature $S(\xbf_i)$ at point $\xbf_i$.}
    \label{fig: Diagram}
\end{figure*}

Our quantum algorithm to find the intrinsic dimension and the noise-resistant scalar curvature is a translation of the algorithm proposed in~\cite{hickok2023intrinsic} into the quantum setting. In our method, we utilize many of the recipes from the recently introduced block-encoding/quantum singular value transformation framework~\cite{gilyen2019quantum, low2017optimal, low2019hamiltonian}. We refer readers to the Appendix~\ref{sec: summaryofnecessarytechniques} for an overview of essential concepts and related recipes.

Let $X = \{\xbf_i\}^N_{i=1} \subset \Rbb^m$, where every vector $\xbf_i \in \Rbb^m$ is a data point in the point cloud $X$. The computational model/assumption of our work consists of:
\begin{enumerate}
    \item Classical knowledge/description of $\{\xbf_i\}^N_{i=1} $. 
    \item Classical knowledge/description of the Euclidean separations $\{ d(\xbf_i,\xbf_j)\}_{i,j=1}^N$ between all pairs of data points, i.e.
    \begin{equation}
        d(\xbf_i,\xbf_j) = \left\| \xbf_i - \xbf_j \right\| \ .
    \end{equation}
    Note that here we only consider an Euclidean embedding space for simplicity, but in general our approach can be applied to any metric space.
\end{enumerate}
The knowledge/description of the classical data in the above assumptions can be understood from the perspective of state preparation \cite{grover2000synthesis,grover2002creating,plesch2011quantum, schuld2018supervised, nakaji2022approximate,marin2023quantum,zoufal2019quantum,zhang2022quantum}. These works address the problem of preparing a state $\sum_i a_i \ket{i}$, provided that the amplitudes $\{a_i\}$ are classical known. Our quantum algorithm would use these state preparation techniques to obtain quantum states $\varpropto \xbf_i$, $\varpropto \sum_{i,j} d(\xbf_i,\xbf_j) \ket{j}$ (see Appendix \ref{sec: blockencodingkernelmatrix}-\ref{sec: summaryofnecessarytechniques}). The efficient---and also, optimal---state preparation protocol proposed in \cite{zhang2022quantum} is an important ingredient that contributes to \textit{quantum speed-up} of our algorithm for estimating local dimension and curvature.

On a given Riemannian manifold, the notion of distance depends on the metric (see Appendix~\ref{sec: reviewofdifferentialgeomtry}), and the geodesic distance between to points is defined as the length of the shortest path connecting them. As we explain in Appendix~\ref{sec: pipelineclassicalalgorithm}, both the underlying manifold of the data point-cloud $X$ and the geodesic distances between points $\{d_G(\xbf_i,\xbf_j)\}^N_{i,j=1}$ from the ``raw'' Euclidean distances $\{d(\xbf_i,\xbf_j)\}^N_{i,j=1}$ can be approximated using the method of diffusion geometry introduced in~\cite{coifman2006diffusion}. From the distance $d (\xbf_i, \xbf_j)$, we define the \textit{affinity kernel} matrix $K$ with the following entries:
\begin{align}
    K_{ij} = \exp\left[- \frac{d^2(\xbf_i, \xbf_j)}{\sigma^2}  \right] \ ,
\end{align}
where $\sigma$ is a \textit{hyperparameter} for the affinity kernel-scale. Let $\{ \lambda_k, \ket{\psi_k}\}$ be the eigenvalues and their corresponding normalized eigenvectors of the matrix $K$. We can estimate the geodesic distance on the underlying manifold between two points $\xbf_i, \xbf_j$ with a \textit{single-timestep} diffusion distance approximation:
\begin{align}
    d^2_G(\xbf_i,\xbf_j) \approx \sum_{k=1}^N \lambda_k^{2t}\Big|_{t=1} \left( \ket{\psi_k}_i - \ket{\psi_k}_j \right)^2 \ ,
\end{align}
where $\ket{\psi_k}_i $ refers to the $i$-th component of the $k$-th eigenvector $\ket{\psi_k}$ (see Appendix \ref{sec: pipelineclassicalalgorithm}). 

The above estimation for geodesic distances by using the spectrum of the affinity kernel matrix $K$ allows us to use the \textit{recently introduced} block-encoding framework \cite{gilyen2019quantum, low2017optimal,low2019hamiltonian}. While a more detailed summary of this framework is given in the Appendix \ref{sec: summaryofnecessarytechniques}, let us explain a few main concepts. A unitary $U$ is said to be a block encoding of $A$ (with operator norm $|A| \leq 1$) if $U$ contains $A$ in the top left corner, i.e.
$$U = \begin{pmatrix}
        A & * \\
        * & * 
    \end{pmatrix} \ , \ $$
    where $(*)$ refers to possibly non-zero entries. Suppose that $U_1$ is a block encoding of $A_1$, $U_2$ is a block encoding of $A_2$, then for some known $\alpha, \beta \leq 1$, we can construct another unitary a unitary block encoding of $\alpha_1 A_1, \alpha_2 A_2$ (Lemma \ref{lemma: scale}) $ \alpha A_1 + \beta A_2$ (Lemma \ref{lemma: sumencoding}, Linear combination), of $ A_1 A_2$ (Lemma \ref{lemma: product}, Multiplication), and also of $A_1 \otimes A_2$ (Ref. \cite{camps2020approximate}, Tensor product). Additionally, for a factor $\gamma > 1$ and with a guarantee $\gamma A \leq \frac{1}{2}$, it is possible to construct the block encoding of $\gamma A$ (Lemma \ref{lemma: amp_amp}, Amplification). In particular, as a central result, given the unitary block-encoding $U$ of a (suppose to be Hermitian for simplicity) matrix $A = \sum_{k} \lambda_k \ket{\psi_k}\bra{\psi_k}$, then there is a constructable quantum circuit that returns the block-encoding of $P(A) =  \sum_{k} P(\lambda_k) \ket{\psi_k}\bra{\psi_k}$, where $P(A)$ is generally a polynomial of bounded norm. As we can see, this framework is naturally suited to handle and perform arithmetic operations on the spectrum of any block-encoded operator. Subsequently in Appendix \ref{sec: blockencodingkernelmatrix}, we will show that from the classical knowledge of pairwise distances $\{d(\xbf_i,\xbf_j)\}_{i,j=1}^N$, it is possible to obtain the block-encoding of $K^\dagger K$. Then, we can leverage the block-encoding arithmetic operations to obtain the block-encoding of an diagonal operator that contains the geodesic distances as entries (see Appendix~\ref{sec: detailgeodesicdistance}).

For a given point $\xbf_i$, let $N_i$ denote the set of its nearest points (in terms of geodesic distance); the size of this neighborhood is $|N_i| = \mathcal{N}$. The so-called centroid is defined as follows,
\begin{align}
        \widetilde{\xbf}_i = \frac{1}{\mathcal{N}} \sum_{j, \xbf_j \in N_i } \xbf_j .
    \end{align}
Earlier, we pointed out that the block-encoding framework can be applied to obtain an (block-encoded) operator containing the geodesic distances on the diagonal. A diagonal operator has a simple, yet very useful property that its eigenvectors are the computational basis states and the corresponding eigenvalues are exactly those entries on the diagonal. Thus, if we want to find the set $N_i$ nearest points to $\xbf_i$, which is equivalent to identify the set of $\mathcal{N}$ smallest geodesic distances in $\{ d_G(\xbf_i,\xbf_j\}_{j=1}^N$, we can first invert this diagonal operator and then find the largest eigenvalues/eigenvectors of the resultant operator. As will be discussed in the Appendix \ref{sec: findingninearestpoints}, this whole procedure can be executed by combining Lemma \ref{lemma: negative} and the \textit{recent development} in quantum PCA  \cite{nghiem2025refined, lloyd2014quantum} (e.g.,  Lemma \ref{lemma: largesteigenvalues}) to find the largest eigenvalues/eigenvectors. As a result, it gives us information about the $\mathcal{N}$ closest points to the data point $\xbf_i$ of interest.

 The centered nearest points to $\xbf_i$ is defined as:
    \begin{align}
        \widetilde{N}_i = \{ \xbf_j- \widetilde{\xbf}_i\}_{j, \xbf_j \in N_i},
    \end{align}
Define the matrix $C_i$ to be the matrix of size $\mathcal{N} \times m$, where the rows of $C_i$ correspond to $\widetilde{N}_i$. The local dimension $d_i$ is defined in the neighborhood of $\xbf_i$ as follows:
  \begin{align}
        d_i = \arg \min_p \left\{ p \ \Bigg|\  \frac{\sum_{\alpha=1}^p \sigma_\alpha^2 }{\sum_{\alpha=1}^{\mathcal{N}} \sigma_\alpha^2} \geq \tau  \right\},
    \end{align}
where $\{\sigma_{\alpha}\}_{\alpha =1}^{\mathcal{N}}$ are the singular values of $C_i$ (assumed to be in descending order $\sigma_1 \geq \sigma_2 \geq ... \geq \sigma_{\mathcal{N}}$). As we demonstrate in Appendix \ref{sec: obtainblockencodingofcici}, from the knowledge of $\mathcal{N}$ nearest points found earlier, we can leverage both state preparation \cite{zhang2022quantum} (see Lemma \ref{lemma: stateprepration}) and block-encoding arithmetic recipes again to obtain the block-encoding of $\varpropto C_i^\dagger C_i$. Then, by applying the quantum PCA Lemma.~\ref{lemma: largesteigenvalues}, we can find the cut off of the eigenvalues at which the above ratio is reached, e.g., see Appendix \ref{sec: estimatinglocalintrinsicdimension}.

Given the set of local intrinsic dimensions $\{d_i\}_{i=1}^N$, we can assess whether the manifold hypothesis holds for the point cloud $X$. If so—i.e., if $X$ can be well-approximated by a lower-dimensional manifold with global intrinsic dimension $d$—then it becomes possible and meaningful to estimate the curvature at a specific data point $\mathbf{x}_i$. Even if the manifold hypothesis fails, as in cases where $X$ is better modeled as a \textit{union of manifolds} with varying dimensionalities~\cite{brown2022verifying,shinde2024geometric}, assigning local curvature to each point can still remains informative when interpreted through neighborhood geometry. We start by define the \textit{density kernel} associated at every point: 
\begin{align}
     \rho(\xbf_i) \approx \sum_{j, \xbf_j \in N_i} \exp\left[ -  \frac{d_G^2(\xbf_i,\xbf_j)}{h^2}\right] \ , 
\end{align} 
in which $h$ is a \textit{hyperparameter} for the density kernel-scale. A geodesic ball of radius $r$ centered at $\mathbf{x}_i$ is the set of all points whose geodesic distance to $\mathbf{x}_i$ is less than or equal to $r$, i.e.:
\begin{align}
        B_r(\xbf_i) = \{ \xbf_j \in X | d_G(\xbf_i,\xbf_j) \leq r \} \ .
    \end{align}
The volume of this ball can be estimated with:
    \begin{align}
        \rm Vol \left(  B_r(\xbf_i)\right) = \sum_{\xbf_j \in B_r(\xbf_i)} \frac{1}{\rho(\xbf_j)} \ ,
    \end{align}
as we have explained in Eq. \eqref{estimate_volume}. 
From the classical knowledge of those $\{d_G(\xbf_i,\xbf_j)\}_{\xbf_j \in B_r(\xbf_i) }$, one can use classical procedure to compute the sampling density, as well as volumes of any geodesic balls of choice. As will be detailed in the Appendix \ref{sec: quantumalgorithmestimatingdensity} and \ref{sec: fitcurvature}, the quadratic fit $ \rm Vol_{nor} \left(  B_{r}(\xbf_i)\right) \  \rm versus \  r^2$ results in the fit parameter to be:
\begin{align}
     A = \frac{ \sum_{j=1}^{\mathcal{N}}\rm Vol_{nor} \left( B_{r_j}(\xbf_i) \right)/\mathcal{N} }{1+ \sum_{j=1}^{\mathcal{N}} d_G^2(\xbf_i,\xbf_j)/\mathcal{N}}
\end{align}
Appendix \ref{sec: fitcurvature} explains how  we use the state preparation technique (Lemma \ref{lemma: stateprepration}) plus Hadamard test to evaluate the terms $ \sum_{j=1}^{\mathcal{N}}\rm Vol_{nor} \left( B_{r_j}(\xbf_i) \right), \sum_{j=1}^{\mathcal{N}} d_G^2(\xbf_i,\xbf_j)$. Then, the value of $A$ can be estimated.
\begin{table*}
    \centering
    \renewcommand{\arraystretch}{2} 
    \begin{tabular}{|c|c|}
    \hline
       \textbf{Objective}  &  \textbf{Complexity} \\
       \hline
      Obtain the block-encoding of $K^\dagger K$    & $\mathcal{O}\left( \log^2\left(\frac{1}{\epsilon} \right) \log N \right)$ \\
      \hline
      Obtain the block-encoding of $ \varpropto  \sum_{j=1}^N d_G^4(\xbf_i,\xbf_j) \ket{j-1}\bra{j-1}$ & $ \mathcal{O}\left( \log (N) \log^2\left( \frac{1}{\epsilon}\right)  \right) $ \\
      \hline
      Obtain the block-encoding of $ \varpropto  \sum_{j=1}^N \frac{1 }{d_G(\xbf_i,\xbf_j)} \ket{j-1}\bra{j-1}$ & $   \mathcal{O}\left( \log^2\left(\frac{1}{\epsilon} \right) \log (N)\log^2 \left( \frac{N}{ \epsilon}\right) \right) $ \\
      \hline
      Find $N_i$ & $   \mathcal{O}\left( \log^2\left(\frac{1}{\epsilon} \right) \log (N)\log^2 \left( \frac{N}{ \epsilon}\right) \log^{\mathcal{N}}\left( \frac{N}{\epsilon} \right) \left(\frac{1}{\epsilon \Delta^{\mathcal{N}}}\right) \log^{\mathcal{N}}\frac{1}{\epsilon} \right) $ \\
      \hline
      Obtaining the block-encoding of $\varpropto \left( C_i^\dagger C_i\right) $ & $ \mathcal{O}\left( \log (m\mathcal{N})\right) $ \\
      \hline
      Estimating the local dimension $d_i$ &  $\mathcal{O}\left( \log (m\mathcal{N}) \frac{1}{\delta^{d_i} \epsilon } \log^{d_i} \left( \frac{m}{\epsilon} \right) \log \left(\frac{1}{\epsilon}\right)  \right)$ \\
      \hline 
Fit the quadratic curve  & $\mathcal{O}\left( \frac{1}{\epsilon}\log \mathcal{N} \right) $\\
\hline
Estimate the curvature $S(\xbf_i)$ & $  \mathcal{O}\left(  \frac{1}{\Delta^{\mathcal{N}} \epsilon} \log^{\mathcal{N}+3} \left( N\right) + \frac{1}{\delta^{d_i} \epsilon } \log\left(m \mathcal{N} \right) \log^{\lceil d_i \rceil} m  \right)$ \\
\hline
    \end{tabular}
    \caption{\textbf{Table summarizing the complexity of the procedure in Fig.~\ref{fig: Diagram}.} $\epsilon$ is the precision parameter. $\Delta$ is defined as following: let $\{ d_G(\xbf_i,\xbf_j) \}_{j=1}^N$ be the set of geodesic distances from $\xbf_i$; sort this set from lowest to highest values; then $\Delta$ is defined as the minimum of the separation between two consecutive (sorted) values. $\delta$ is defined as following: let $\sigma_1 \geq \sigma_2 \geq ... \geq \sigma_{\mathcal{N}}$ be the singular values of $C_i$, then $\delta \equiv \min \{  | \sigma_{i}-\sigma_{i+1}|  \}_{i=1}^{d_i}$. The derivation for these reported estimations can be found in Appendix \ref{sec: complexityanalysis}.}
    \label{tab: summarizecomplexity}
\end{table*}
The pipeline of our quantum algorithm for estimating the local intrinsic dimension $d_i$ and curvature $S(r_i)$ at the data point $\xbf_i$ is summarized in Fig.~\ref{fig: Diagram}. While a detailed analysis of its complexity will be given in the Appendix \ref{sec: complexityanalysis}, we refer to Table \ref{tab: summarizecomplexity} for a complexity list of all the steps appeared in Fig.~\ref{fig: Diagram}. We summarize our main result in the following.
\begin{theorem}
\label{thm: 1}
    Provided the data points $X = \{ \xbf_1,\xbf_2,...,\xbf_N \} $ and classical value of distances $\{ d(\xbf_i, \xbf_j)\}_{i,j=1}^N$ between all pairs of data points are given. Let $N_i$ be the (local) neighborhood of $\xbf_i$ with size $|N_i| = \mathcal{N}$. The quantum algorithm in Fig.~\ref{fig: Diagram}, as assisted by a classical algorithm of at most $\mathcal{O}(\log N)$ cost, outputs the estimation of the local dimension $d_i$ and local curvature $S(\xbf_i)$---up to an additive accuracy $\epsilon$---with complexity\footnote{
    For more detail, we show in Appendix~\ref{sec: complexityanalysis} that this complexity is about
    $$
        \mathcal{O}\left(  \frac{1}{\Delta^{\mathcal{N}} \epsilon}\log^{\mathcal{N}+3} \left( \frac{N}{\epsilon}\right) + \frac{1}{\delta^{d_i} \epsilon}\log\left(m \mathcal{N} \right) \log^{d_i} m  \right) \ , $$
where $\Delta, \delta$ are some constants depending on the dataset $X$.}:
    \begin{equation}
    \begin{split}
        \mathcal{O}\left( \frac{1}{\epsilon}\log^{\mathcal{N}+3} \frac{N}{\epsilon} \right) \ \ &\text{when $N\gg m$} \ ,
        \\
        \mathcal{O}\left(\frac1\epsilon \log^{d_i+1} m \right) \ \ &\text{when $m \gg N$} \ .
    \end{split}
    \label{complexity_est}
    \end{equation}  
\end{theorem}

\section{Discussion}
\label{sec: discussion}
In the following, we discuss our results from a broader perspective. We particularly show its potential advantage compared to the classical counterpart and discuss a few corollaries and possible extension of our technique toward related computational problems. \\

\noindent
\textbf{Classical algorithm \cite{hickok2023intrinsic}.} The classical algorithm's computational cost mainly comes from Step 1 (computing geodesic distances) and Step 2 (finding local dimension). As we approximate the geodesic distances via the spectrum of the affinity kernel matrix $K$, we first need to (classically) compute the entries of $K$. As the computation involves evaluating all affinity pairs $K_{ij}$ for $i,j=1,2,...,N$, this classical procedure has complexity $\mathcal{O}( N^2)$. Next, we need to perform the exact diagonalization on $K$ to find its eigenvectors/eigenvalues, which incurs a complexity $\mathcal{O}\left( N^3 \right)$, as the matrix $K$ is of size $N \times N$. Next, we need to evaluate all the geodesic distances $\{ d_G(\xbf_i,\xbf_j)\}_{j=1}^N$, which takes further complexity $\mathcal{O}\left( N\right)$. As the next step, we need to build the matrix $C_i$, which is of dimension $\mathcal{N}\times m$. Performing singular value transformation on this matrix would take complexity $\mathcal{O}\left( \max (\mathcal{N},m)^3 \right) = \mathcal{O}(m^3)$, which is typically the case for high-dimensional data ($m>\mathcal{N}$). Finally, the computation of geodesic balls volume and performing quadratic fitting have complexity $\mathcal{O}\left( \mathcal{N} \right)$, negligible compared to other contributions. Thus, after summing up, the total classical complexity is $\mathcal{O}\left( N^3 \right)$. \\

\noindent
\textbf{Potential quantum advantage.} From Eq. \eqref{complexity_est}, our quantum algorithm has polylogarithmic scaling in both $N$ and $m$, offering an \textit{exponential speed-up} compared to the above classical algorithm. Interestingly, the degree of speed-up also depends on the local dimension $d_i$. This is quite analogous to existing quantum topological data analysis algorithms~\cite{lloyd2016quantum, schmidhuber2022complexity, berry2024analyzing, nghiem2023quantum, lee2025new}, where the quantum speedup in estimating Betti numbers to some multiplicative accuracy also depends on the Betti numbers themselves, e.g., quantum algorithms perform better in the regime where the simplicial complex of interest exhibits high Betti numbers, or that the topological space has many ``holes''. In our case, if the data points tend to ``live'' on a low-dimensional manifold, then it is the regime where our quantum algorithm performs most efficiently. \\

\noindent
\textbf{Application.} Earlier, we have introduced the affinity kernel matrix with the entries $ K_{ij}$, which helps approximating the geodesic distance on the underlying manifold of data point-cloud $X$. This matrix also turns out to be common in the context of the diffusion map \cite{coifman2006diffusion}. In this context, one desires to obtain a low-dimensional representation of the given data points. A more detailed description can be found in Appendix \ref{sec: diffusionmap}. Here, we point out that, from the kernel matrix $K$, one can build the so-called diffusion operator $P$. By performing a spectral decomposition on $P$, one obtains the low-dimensional representation of the given, say $\xbf_i$, as the $i$-th components of the top $n$ eigenvectors, multiplied by a power of the corresponding eigenvalues. The value of $n$ controls the dimension of the space that we wish to project onto, and in practice, it is usually $2$ or $3$. 

Our quantum algorithm for diffusion map is a straightforward corollary of the procedure underlying diagram~\ref{fig: Diagram}. We defer the full description of the quantum algorithm to part 2 of Appendix~\ref{sec: diffusionmap}, and recapitulate the result in the following theorem.
\begin{theorem}
\label{thm: 2}
     Provided the data points $X = \{ \xbf_1,\xbf_2,...,\xbf_N \} \subseteq \Rbb^m$ and classical value of distances $\{ d(\xbf_i, \xbf_j)\}_{i,j=1}^N$ between all pair of data points. Then for a given data point $\xbf_i$, there is a quantum algorithm that estimates $n$ entries of its $n$-dimensional (with $n \ll m $) representation. For an estimation of additive accuracy $\epsilon$, the algorithm has complexity
   \begin{align}
      \mathcal{O}\left(  \log^{n+1} (N)+  \log^{2n+6}  \frac{1}{\epsilon}  \right).
\end{align}
\end{theorem}
We point out that, previously, there has been an attempt to develop a quantum algorithm for diffusion map~\cite{sornsaeng2021quantum}. Their method requires oracle access to certain matrices, with a total running time $\mathcal{O}\left( N^2 \log^3 N   \right)$. In comparison to this work, ours does not require oracle access; rather, we only need the classical values of the pairwise distances $\{d(\xbf_i,\xbf_j)\}_{i,j=1}^N$. Additionally, for $n= \mathcal{O}(1)$ as we pointed out earlier, our method's complexity yields almost an exponential speedup compared to~\cite{sornsaeng2021quantum}.

\section{Conclusion}
\label{sec: conclusion}
In this work, we have further explored the potential of quantum computers towards GDA. This is a new, rapidly growing field that borrows techniques from modern geometry theory to analyze large-scale datasets. We have specifically focused on three problems: local dimension, local curvature, and local/low-dimensional representation. We have shown that under appropriate assumptions, quantum computers can estimate the intrinsic dimension and local curvature exponentially faster than their classical counterparts. Building on this, we extend the related technique to the context of diffusion maps and demonstrate that quantum computers can also compute the low-dimensional features of data points, provided they are given the appropriate information from the original, higher-dimensional space. 

As mentioned in the introduction, a few efforts have been made to investigate the capabilities of quantum algorithms in the field of topological data analysis. Despite some interesting results having been obtained, the complexity-hardness established in~\cite{schmidhuber2022complexity} has placed a barrier on the extent to which quantum advantage can actually be gained. At the same time, our work has suggested that GDA is a promising avenue for exploring quantum computational advantage. This is a relatively new field that remains largely unexplored, particularly from a quantum perspective. We therefore believe that the demonstration of quantum speedups in this work can serve as a great motivation for future study. For example, in the context of Theorems~\ref{thm: 1} and~\ref{thm: 2}, there is an important assumption that the dataset $X$ belongs to some manifold, i.e., the manifold hypothesis. Whether our algorithm, and the corresponding classical algorithm, can be applied beyond the manifold hypothesis (and, to a certain extend, the union of manifolds~\cite{brown2022verifying,shinde2024geometric}), is an interesting future question.

\section*{Acknowledgements}
N.A.N. acknowledges support from the Center for Distributed Quantum Processing. Parts of this work were completed while N.A.N. is an intern at QuEra Computing Inc. N.A.N. and T.W. acknowledge the support by the Center for Distributed
Quantum Processing at Stony Brook University. T.V.P. would like to thank Truong H. Cai for useful discussions.

\bibliography{ref.bib}
\bibliographystyle{unsrt}

\appendix
\onecolumngrid

\section{A Derivation for the Geodesic-Ball Volume-Formula}
\label{sec:ball_formula}

At any point $p \in \mathcal{M}$, the manifold is locally flat, so one can introduce Riemann normal-coordinates $x \in \mathbb{R}^d$ centered at $p$, in which the metric tensor represented in those coordinates satisfies $g_{\mu\nu}(0)=\delta_{\mu\nu}$ (thus $g(0)$ is the identity matrix $\mathbb{I}$) and $\partial_\rho g_{\mu\nu}(0)=0$, and and the geodesics through $p$ appear as straight coordinate lines. At the nearby neighborhood, the metric admits the following expansion \cite{spivak1979volume2,sternberg2013curvature,hatzinikitas2000note}:
\begin{equation}
    g_{\mu\nu}(x) \approx \delta_{\mu\nu} - \frac13 R_{\mu\rho\nu\sigma}(0) x^\rho x^\sigma + \mathcal{O}(x^4) \ ,
\end{equation}
and therefore the determinant can be computed using the standard identity for small symmetric perturbations of the identity matrix $\mathbb{I}$ ($\det(\mathbb{I}+\epsilon A)\approx 1+\epsilon\cdot\mathrm{trace}(A))$:
\begin{equation}
    \text{det}\left[ g(x) \right] \approx 1 - \text{Tr}\left[ \frac13 R_{\mu\rho\nu\sigma}(0) x^\rho x^\sigma \right] + \mathcal{O}(x^4) = 1 - \frac13 C_{\rho \sigma}(0) x^\rho x^\sigma + \mathcal{O}(x^4) \ ,
\end{equation}
where $R_{\mu\rho\nu\sigma}(0)$ and $C_{\rho\sigma}(0)$  denote the Riemann curvature tensor and the Ricci curvature tensor respectively evaluated at point $p$, see Appendix \ref{sec: reviewofdifferentialgeomtry}.

A geodesic ball $B_r(p)$ of radius $r$ centered at $p$ consists of all points whose geodesic distance from $p$ is less than or equal to $r$. Since the geodesics align with straight lines in Riemann normal coordinates, the geodesic ball corresponds to the set of all points satisfy $|x| <r$. The volume of the ball $B_r(p)$, with respect to the Riemannian metric $g$, is calculated with:
\begin{equation}
\begin{split}
    \text{Vol}[B_r(p)] &= \int_{|x|<r} d^d x \sqrt{\text{det}\left[ g(x) \right]} \approx \int_{|x|<r} d^d x \left[ 1 - \frac13 C_{\rho \sigma}(0) x^\rho x^\sigma + \mathcal{O}(x^4)  \right]^{1/2}
    \\
    & \approx \int_{|x|<r} d^d x \left[ 1 - \frac16 C_{\rho \sigma}(0) x^\rho x^\sigma + \mathcal{O}(x^4)  \right] = \omega_d r^d \left[ 1 - \frac{S(0)}{6(d+2)} + \mathcal{O}(r^4) \right] \ ,
\end{split}
\end{equation}
where $\omega_d = \pi^{d/2}/\Gamma(d/2+1)$ and $S(0)=\text{Tr}[C_{\rho\sigma}(0)]$ is the scalar curvature evaluated at point $p$. Higher-order terms in the ball-volume expansion with respect to the geodesic radius $r$ can be found in \cite{gray1979riemannian}.

\section{A pipeline for classical algorithm}
\label{sec: pipelineclassicalalgorithm}

In this section we provide a detailed description of the classical algorithm for estimating the curvature of a point cloud, which was introduced in \cite{hickok2023intrinsic}. The algorithm contains 6 main steps, and in the following, we will describe them one by one. \\

\begin{method}[Classical Algorithm for Estimating (Local) Curvature]
Let $X = \{\xbf_1, \xbf_2, ..., \xbf_N\} \subset \Rbb^m$ where $\xbf_i \in \Rbb^m$ be the set of data points, or point-cloud, and pairwise distances $d_{ij} \equiv d({\bf x}_i,{\bf x}_j)$ among them are given.
\end{method}
\noindent
\textbf{Step 1: Estimate Geodesic Distances.} \\
\noindent
To estimate the geodesic distance between points ${\bf x}_i$ and ${\bf x}_j$ via diffusion geometry \cite{coifman2006diffusion}, one proceeds in six main steps:
\begin{itemize}
    \item Build a smooth affinity matrix where the kernel‐scale $\varepsilon$ sets the (Euclidean) size of influence between points:
    \begin{equation}
        K_{ij} \equiv \exp(-d_{ij}^2/\sigma^2) \ .
    \end{equation}
    A popular choice for the \textit{hyperparameter} $\sigma^2$ is the median of all $d^2_{ij}$ (excluding the diagonals $d_{ii}=0$), but the optimal value is often problem-dependent.
    \item Normalize the affinity matrix to a diffusion operator:
    \begin{equation}
        P_{ij} = K_{ij}/\sum_k K_{ik} \ \ \text{so that} \ \ \sum_j P_{ij}= 1 \ ,
    \end{equation}
    and $P$ represents one step of a random walk on the data point-cloud $X$ \cite{coifman2006diffusion}.
    \item Find the eigenvalues $\{ \lambda_k\}$ and right (normalized) eigenvectors $\{\psi_k\}$ of the operator $P$. In other words, we do a spectral decomposition for $P$, i.e.:
    \begin{equation}
        P_{ij} = \sum^N_{k=1} \lambda_k \psi_{ik} \psi_{jk} \ .
    \end{equation}
    where $\psi_{ik}$ is the $i$-th component of the eigenvector $\psi_k$.
    \item  In the continuum limit (dense-sampling), $P$ approximates the Laplace–Beltrami operator. The $t$-timestep diffusion distance between points ${\bf x}_i$ and ${\bf x}_j$, denoted as $\mathcal{D}^2_{t}({\bf x}_i,{\bf x}_j)$, is given by: 
    \begin{equation}
        \mathcal{D}_{t}({\bf x}_i,{\bf x}_j) =\left[  \sum^N_{k=1} \lambda_k^{2t}(\psi_{ik} - \psi_{jk})^2 \right]^{1/2} \ .
    \end{equation}
    We can define the geodesic distance by $d_G(\xbf_i,\xbf_j) \approx \mathcal{D}_{t}({\bf x}_i,{\bf x}_j)$, which follows from the Varadhan’s asymptotic formula \cite{varadhan1966asymptotic}:
    \begin{equation}
        \lim_{t \rightarrow 0} \mathcal{D}_{t}({\bf x}_i,{\bf x}_j) \propto d_G({\bf x}_i,{\bf x}_j) \ .
    \end{equation}
    A good choice for the \textit{hyperparameter} $t>0$ (does not have to be an integer) must be large enough to model the \textit{local continuity} between discrete points, but should also be small enough so that this approximation is a good estimate for the length of the geodesic path on the underlying manifold embedded in $\mathbb{R}^m$ \cite{coifman2006diffusion}. For simplicity, in this work we do the geodesic estimate with a \textit{single-timestep} $t=1$.
\end{itemize}

\ \ 

\noindent
\textbf{Step 2: Estimate the intrinsic dimension $d$.} \\
\noindent
Define $d_G$ as the geodesic distance matrix, which is of size $N \times N$ and the entry $(i,j)$ contains the geodesic distance $d_G(\xbf_i, \xbf_j)$ between the point $\xbf_i$ and $\xbf_j$. 
\begin{itemize}
    \item At each point, say $\xbf_i$, choose a local neighborhood of a \textit{fixed number} $\mathcal{N}$ nearest points in $d_G$ (note that now the definition of nearest refer to the geodesic distance instead of the Euclidean distance as in the previous step). Let $N_i$ denotes the set of those nearest points of $\xbf_i$, then $|N_i|=\mathcal{N}$. Typically, we select $N \gg \mathcal{N}$.
    \item Define the \textit{center} of the neighborhood chosen at $\xbf_i$ as:
    \begin{align}
        \widetilde{\xbf}_i = \frac{1}{|N_i|} \sum_{j, \xbf_j \in N_i } \xbf_j
    \end{align}
    \item Define the \textit{centered-coordinate} of the neighborhood chosen at $\xbf_i$ as:
    \begin{align}
        \widetilde{N}_i = \{ \xbf_j- \widetilde{\xbf}_i\}_{j, \xbf_j \in N_i}
    \end{align}
    \item Define the matrix $C_i$ to be the matrix of size $|N_i| \times m$, where the rows of $C_i$ is in correspondence with $\widetilde{N}_i$.
    \item Perform singular decomposition on $C_i$ and obtain a series of singular values, i.e.
    \begin{equation}
        C_i = \mathbb{U} \mathbb{D} \mathbb{V}^{\top} \ , \ \text{where} \  \mathbb{D}_i = \begin{pmatrix} \sigma_1 & 0 & 0 & ... \\ 0 & \sigma_2 & 0 & ...\\  0 & 0 & \sigma_3 & ... \\ ... & ... & ... & ... \end{pmatrix} \ \text{and} \ \mathbb{U} = (\vec{\mathbb{U}}_1, \vec{\mathbb{U}}_2, \vec{\mathbb{U}}_3, ...) \ .
    \end{equation}
    The set of orthogonal vectors $\{ \vec{\mathbb{U}}_\alpha \}$ ($\alpha=1,2,...,|N_i|$) are the eigenvectors of the symmetric matrix $C_i C_i^\top$, corresponding to the eigenvalues $\{ \sigma_\alpha \}$ that are conventionally ordered $\sigma_1 \geq \sigma_2 \geq ... \geq \sigma_{|N_i|}$. 
    
    Each $\sigma_\alpha$ is the \textit{singular value} of the point-cloud along the corresponding principal direction $\vec{\mathbb{U}}_\alpha$. Specifically, $\lambda_\alpha \equiv \sigma^2_\alpha |N_i|^{-1/2}$ represents the variance of the data captured by $\vec{\mathbb{U}}_\alpha$.
    \item Estimate the local dimension $d_i$ in the neighborhood of $\xbf_i$ by finding the integer $p$ so that:
    \begin{align}
        d_i = \arg \min_p \left\{ p \ \Bigg|\  \frac{\sum_{\alpha=1}^p \lambda_\alpha }{\sum_{\alpha=1}^{|N_i|} \lambda_\alpha} = \frac{\sum_{\alpha=1}^p \sigma_\alpha^2 }{\sum_{\alpha=1}^{|N_i|} \sigma_\alpha^2} \geq \tau  \right\}
    \end{align}
    where $\tau \in (0.9,0.99)$ is the threshold. In other words, we require that the subspace spanned by $\{\vec{\mathbb{U}}_\alpha\}^{d_i}_{\alpha=1}$ to explain at least a fraction $\tau$ of the point-cloud's total variance.
    \item Repeat the above procedure for all data points, we obtain a set of local dimensions $d_1,d_2,...,d_N$ 
    \item Find the global estimation of intrinsic dimension $d$ as the \textit{median} of $ \{ d_1,d_2,...,d_N \}$. Although the neighborhood's size $\mathcal{N}$ is selected beforehand (in \textbf{Step 1}), one should check that $\mathcal{N} > d$ post hoc to ensure each local neighborhood has contained enough points to reliably estimate the $d$-dimensional tangent space and separate intrinsic structures from noise.
\end{itemize}

\noindent
\textbf{Step 3: Density estimation.} \\
\noindent
The so-called sampling density $\rho(\xbf_i)$ at the data point $\xbf_i$ is estimated as follows:
\begin{align}
    \rho(\xbf_i) \approx \sum_{j, \xbf_j \in N_i} \exp\left[ - \left( \frac{d_G(\xbf_i,\xbf_j)}{h}\right)^2 \right] \equiv \sum_{j, \xbf_j \in N_i} w_{ij} \ ,
\end{align}
where $h$ is a global scale parameter which controls the locality and $d_G$ is the geodesic distance from Step 1. A good choice for $h$ is the typical geodesic distance between points in a neighborhood. Note that, here, the Gaussian weight
\begin{equation}
w_{ij} = \exp\left[ - \left( \frac{d_G(\xbf_i,\xbf_j)}{h}\right)^2 \right]
\end{equation}
is exactly the heat kernel (i.e. the fundamental solution of the diffusion equation at ``time'' $ \propto h^2$), so summing these weights furnishes a diffusion‑smoothed estimate of the local sampling density. \\

\noindent
\textbf{Step 4: Volume estimation of geodesic balls.} 

\noindent
\begin{itemize}
    \item Choose a range of radii $r_1,r_2, ..., r_M$ within some known range $[r_{\min}, r_{\max}]$.  
    \item For a radii $r$ and a data point $\xbf_i$, define the geodesic ball around $\xbf_i$ as follows:
    \begin{align}
        B_r(\xbf_i) = \{ \xbf_j \in X | d_G(\xbf_i,\xbf_j) \leq r \}
    \end{align}
    \item Estimate the volume of the geodesic ball using an inverse-density Monte-Carlo estimator:
    \begin{align}
        \rm Vol \left(  B_r(\xbf_i)\right) = \sum_{\xbf_j \in B_r(\xbf_i)} \frac{1}{\rho(\xbf_j)} \ ,
    \end{align}
    so that points in sparsely sampled regions contribute larger volume estimates (while points in densely sampled regions contribute smaller ones).
    \item Normalize the above volume with the volume of a unit ball $w_d$ in $\Rbb^d$ scaled by a radius $r$:
    \begin{align}
        \rm Vol_{nor} \left(  B_r(\xbf_i)\right) &= \frac{\rm Vol \left(  B_r(\xbf_i)\right) }{ w_d r^d} \\
       \text{where \ }  w_d &= \frac{\pi^{d/2}}{\Gamma (\frac{d}{2} + 1 )}
    \end{align}
\end{itemize}

\noindent
\textbf{Step 5: Fit quadratic of volume versus radius.} \\
\noindent
Recall that at the first step of Step 4 above, we choose a range of radii $r_1,r_2,..,r_M$ and find their corresponding volume of the geodesic ball $ \rm Vol_{nor} \left(  B_{r_1}(\xbf_i)\right),  \rm Vol_{nor} \left(  B_{r_2}(\xbf_i)\right), ...,  \rm Vol_{nor} \left(  B_{r_M}(\xbf_i)\right)   $. From these data, we perform the quadratic fit:
\begin{align}
     \rm Vol_{nor} \left(  B_{r}(\xbf_i)\right) \  \rm versus \  r^2
\end{align}
from which the value of the curvature $S(\xbf_i)$ can be inferred. More specifically, we choose the fit model as:
\begin{align}
     \rm Vol_{nor} \left(  B_r(\xbf_i)\right)  = 1 + A r^2
\end{align}
where $A$ is the scalar parameter of interest. We minimize the following cost function:
\begin{align}
    C = \sum_{j=1}^M \left\| 1+ A r_j^2 -  \rm Vol_{nor} \left(  B_{r_j}(\xbf_i)\right) \right\|^2
\end{align}

\noindent
\textbf{Step 6: Output scalar curvature estimates.} \\
\noindent
The value of $A$ obtained from the quadratic fit above is approximately 
\begin{align}
    A \approx \frac{-S(\xbf_i)}{6(d+2)}
\end{align}
The value of curvature at $\xbf_i$ is then:
\begin{align}
    S(\xbf_i) = -6 (d+2) A
\end{align}

\section{Quantum algorithm for block-encoding the kernel matrix $K$}
\label{sec: blockencodingkernelmatrix}
\noindent
We first mention the following technique introduced in \cite{nghiem2025refined}:
\begin{lemma}
\label{lemma: pairwisesquare}
    Let $U$ (assumed to have depth $T_x$) be some unitary of dimension $> N \times N$ that contains a  vector $\xbf = \sum_{i=1}^N x_i \ket{i-1}$ as the first column. Then there is a quantum circuit of depth $\mathcal{O}\left( T_x+  \rm log (N)\right)$ which is a block encoding of a matrix that contains $\sum_{i=1}^N x_i^2 \ket{i-1}$ as the first column.
\end{lemma}
\noindent
For completeness, we directly quote their proof as follows. \\

\noindent
\textbf{Proof of Lemma \ref{lemma: pairwisesquare}.} First, we consider the matrix of size $N \times N$ on the top-left corner of $U$. Denote this matrix as $A_x$, and the first column of this matrix is $\xbf$. It can also be seen that $U_x$ is the block encoding of $A_x$. Then we can use \ref{lemma: tensorproduct} to construct the block encoding of $A_x \otimes A_x$. The first column of this matrix is $\xbf \otimes \xbf$, which is:
\begin{align}
    \xbf \otimes \xbf &= \sum_{i=1}^n x_i \ket{i-1} \otimes \sum_{i=1}^n x_i \ket{i-1} \\
    &=\sum_{i,j=1}^n x_i x_j \ket{i-1}\ket{j-1} \\
    &= \sum_{i=1}^n x_i^2 \ket{i-1}\ket{i-1} + \sum_{i \neq j} x_i x_j \ket{i-1}\ket{j-1}
\end{align}
Next, we use the permutation:
\begin{lemma}[\cite{barenco1995elementary, shende2005synthesis}]
\label{lemma: permutation}
    Let $\mathcal{H}$ be some $N$-dimensional Hilbert space and $\{ \ket{0}, \ket{1}, \ket{2}, ..., \ket{N} \}$ are basis. Then for a known permutation of basis $ \{ \ket{i-1} \leftrightarrow \ket{j}\} $, there exists a permutation unitary circuit $U_{\rm permutation}$ of depth $\mathcal{O}\left( \log (N \log N)\right)$.
\end{lemma}
\noindent
that achieve the following permutation:
\begin{align}
    \ket{0} \ket{0} &\leftrightarrow \ket{0}\ket{0}\\
    \ket{0} \ket{1} &\leftrightarrow  \ket{1}\ket{1}\\
    &\vdots \\
    \ket{0}\ket{n-1} &\leftrightarrow \ket{n-1}\ket{n-1}
\end{align}
and the remaining basis permutation can be arbitrary. We use such permutation unitary $U_{\rm permute}$ and \ref{lemma: product} to construct the block encoding of $U_{\rm permute} (A_x \otimes A_x)$. The first column of this matrix is:
\begin{align}
    U_{\rm permute} (\xbf \otimes \xbf) &= U_{\rm permute}  \sum_{i=1}^n x_i^2 \ket{i-1}\ket{i-1} + \sum_{i \neq j} x_i x_j \ket{i-1}\ket{j-1}\\
    &= \sum_{i=1}^n x_i^2 \ket{0}\ket{i-1} + (...)
\end{align}    
where $(...)$ refers to the redundant part. The first part is $\sum_{i=1}^n x_i^2 \ket{0}\ket{i-1}  $. Therefore, if we restrict to the top left corner matrix of dimension $n \times n$, then the first column is exactly $ \sum_{i=1}^n x_i^2 \ket{i-1} $.  $\blacksquare$ \\

\noindent
The above procedure can be modified to yield the following more general lemma: 
    \begin{lemma}
\label{lemma: pairwisesquaregeneralized}
    Let $U$ (assumed to have depth $T_x$) be some unitary of dimension $> N \times N$ that contains a  vector $\xbf = \sum_{i=1}^N x_i \ket{i-1}$ as the first column. For some $p \in \mathbb{Z}$, there is a quantum circuit of depth $\mathcal{O}\left( T_x+  p \log (N)\right)$ which is a block encoding of a matrix that contains $\sum_{i=1}^N x_i^p \ket{i-1}$ as the first column.
\end{lemma}

\noindent
The proof of the above lemma, or the modification of the procedure is as follows. Earlier, we consider the state $\xbf\otimes \xbf$ and then use permutation to move those entries $\{ x_i^2 \}_{i=1}^N$. Now we simply need to consider the state $\xbf^{\otimes p}$ and consider the following permutation:
\begin{align}
    \ket{0}^{\otimes p-1} \ket{0} &\leftrightarrow \ket{0}^{\otimes p}\\
    \ket{0}^{\otimes p-1} \ket{1} &\leftrightarrow  \ket{1}^{\otimes p}\\
                        &\vdots \\
    \ket{0}^{\otimes p-1}\ket{n-1} &\leftrightarrow \ket{n-1}^{\otimes p}
\end{align}
with the rest of the basis state permuted arbitrarily. According to \cite{barenco1995elementary, shende2005synthesis}, the above permutation can be obtained with a circuit of complexity $\mathcal{O}\left( p \log N \right)$. $\blacksquare$ \\

Recall that the kernel matrix $K$ is defined as:
\begin{align}
     K_{ij} = \exp(- \frac{d (\xbf_i, \xbf_j)^2}{\sigma^2 } ) 
\end{align}
where $d (\xbf_i, \xbf_j)$ is the pairwise distance between two data points $\xbf_i, \xbf_j$. To proceed, first we recall the following result from \cite{zhang2022quantum}:
\begin{lemma}[Efficient state preparation]
\label{lemma: stateprepration}
    A $n$-dimensional quantum state $\ket{\Phi}$ with known entries (assuming they are normalized to one) can be prepared with a circuit of depth $\mathcal{O}\big( \log (s\log n)\big)$, using $\mathcal{O}(s)$ ancilla qubits ($s$ is the sparsity, or the number of non-zero elements of $\ket{\Phi}$) and a classical pre-processing of complexity $\mathcal{O}( \log n)$. 
\end{lemma}
We remark that while the above state preparation procedure generally requires a classical pre-processing of complexity $\mathcal{O}(\log n)$, it can be improved in certain settings. For example, assume that $\ket{\Phi} = \sum_{i=1}^n a_i \ket{i}$ with $\{a_i\}_{i=1}^n$ known. Then if many of the entries among these $n$ entries are similar, then the result of one classical pre-processing step can be applied to many entries. In this case, the total complexity can be reduced, with the best case being $\mathcal{O}(1)$. 

\noindent
The classical knowledge of pairwise distance $d_{ij} \equiv d( \xbf_i, \xbf_j)$  and the above lemma allow us to obtain the unitary $U_d$ that prepares the following state:
\begin{align}
    \ket{\phi} = \frac{1}{||D||} \sum_{i,j=1}^N \ket{i-1} d_{ij} \ket{j}
\end{align}
where $||D|| = \sqrt{ \sum_{i,j=1}^N d_{ij}^2} $. It is straightforward to see that the first column of the unitary $U_d$ is $\ket{\phi}$. Our goal is to obtain the state (encoded in some matrix) $\varpropto \sum_{i,j=1}^N \ket{i-1} \exp(- \frac{d_{ij}}{\sigma^2} )\ket{j}$ from the unitary $U_d$. To achieve this, we first use Lemma \ref{lemma: pairwisesquaregeneralized} and Lemma \ref{lemma: pairwisesquare} to obtain the block-encodings of matrices having the first columns as:
\begin{align}
     \frac{1}{||D||^p} \sum_{i,j=1}^N \ket{i-1} d_{ij}^p \ket{j},  \frac{1}{||D||^{p-1}} \sum_{i,j=1}^N \ket{i-1} d_{ij}^{p-1} \ket{j}, ...,  \frac{1}{||D||} \sum_{i,j=1}^N \ket{i-1} d_{ij} \ket{j}
     \label{eqn: power}
\end{align}
Next, we point out the following approximation from \cite{trefethen2019approximation}:
\begin{proposition}[\cite{trefethen2019approximation}]
    On any compact interval $[-a,a]$, we have that the Gaussian function $\exp(-x^2)$ is infinitely differentiable and analytic, and that:
    \begin{align}
        \sup_{x \in [-a,a]} | \exp(-x^2)  - P_p(x) | \leq C \exp(-\alpha p).
    \end{align}
    where $p_n(x)$ is the Chebyshev polynomial, which satisfies the following recurrence relation:
    \begin{align}
        P_0(x)= 1, P_1(x) = x\\
        P_{p+1}(x) = 2x P_p(x) - P_{p-1}(x) 
    \end{align}
    \textbf{Fact:} The value of $C$ and $\alpha$ in the above depends on the value of $a$. For $a=1$, then as analyzed in \cite{trefethen2019approximation}, $C \approx 0.1, \alpha \approx 1.09$, which are both $\mathcal{O}(1)$.
\end{proposition}

To apply the above result to our problem, we first need to figure out the value of $p$, which is the degree of the polynomial for approximation. By setting $C \exp(-\alpha p) =\epsilon$, we have that:
\begin{align}
    p = \frac{1}{\alpha} \log \left( \frac{C}{\epsilon}\right) = \mathcal{O}\left( \log \frac{1}{\epsilon}\right)
\end{align}
For convenience, let the polynomial $P_p(x) = \sum_{k=1}^p \alpha_i x^k$. We then use the block-encodings of matrices containing states in Eqn.~\ref{eqn: power} and Lemma \ref{lemma: sumencoding} to construct the unitary block-encoding, denoted as $U_D$, of a matrix, that has the following vector as the first column:
\begin{align}
    \frac{1}{\alpha} \sum_{i,j=1}^N \ket{i-1} P_p\left( \frac{d_{ij}}{||D||} \right) \ket{j} \approx \frac{1}{\alpha}\sum_{i,j=1}^N \ket{i-1} \exp\left( -\frac{d_{ij}^2}{||D||^2} \right) \ket{j} 
\end{align}
where $\alpha = \sqrt{\sum_{k=1}^p \alpha_i^2}$. We note that, if we choose $\sigma = ||D||$, then the entry $\exp\left( -\frac{d_{ij}^2}{||D||^2} \right)   $ is exactly the entry of kernel $K$ defined earlier. On the other hand, if we wish to choose $\sigma \neq ||D||$, then we can slightly modify the above procedure as follows. We use Lemma \ref{lemma: scale} to transform the block-encoded columns in Eqn.~\ref{eqn: power}:
\begin{align}
     \frac{1}{ \alpha ||D||^p } \sum_{i,j=1}^N \ket{i-1} d_{ij}^p \ket{j} \longrightarrow  \frac{1}{ \alpha||D||^p} \sum_{i,j=1}^N \ket{i-1} \frac{d_{ij}^p}{\sigma^p} \ket{j} \\
      \frac{1}{ \alpha||D||^{d-1}} \sum_{i,j=1}^N \ket{i-1} d_{ij}^{p-1} \ket{j} \longrightarrow  \frac{1}{ \alpha||D||^{p}} \sum_{i,j=1}^N \ket{i-1} \frac{d_{ij}^{p-1}}{\sigma^{p-1}} \ket{j} \\
      \vdots \\
      \frac{1}{ \alpha||D||} \sum_{i,j=1}^N \ket{i-1} d_{ij} \ket{j} \longrightarrow \frac{1}{ \alpha||D||^p} \sum_{i,j=1}^N \ket{i-1} \frac{ d_{ij}}{\sigma} \ket{j}
\end{align}
Then we use Lemma \ref{lemma: sumencoding} to construct the block-encoding of a matrix that has the following vector as the first column:
\begin{align}
  \frac{1}{ \alpha||D||^p}  \sum_{i,j=1}^N \ket{i-1} P_p\left( \frac{d_{ij}}{\sigma} \right) \ket{j} \approx  \frac{1}{ \alpha||D||^p}  \sum_{i,j=1}^N \ket{i-1} \exp\left( -\frac{d_{ij}^2}{\sigma^2} \right) \ket{j} 
\end{align}
To obtain the block-encoding of $K$, we point out the following result from \cite{gilyen2019quantum}:
\begin{lemma}[\cite{gilyen2019quantum} Block Encoding Density Matrix]
\label{lemma: improveddme}
Let $\rho = \Tr_A \ket{\Phi}\bra{\Phi}$, where $\rho \in \mathbb{H}_B$, $\ket{\Phi} \in  \mathbb{H}_A \otimes \mathbb{H}_B$. Given unitary $U$ that generates $\ket{\Phi}$ from $\ket{\bf 0}_A \otimes \ket{\bf 0}_B$, then there exists a highly efficient procedure that constructs an exact unitary block encoding of $\rho$ using $U$ and $U^\dagger$ a single time, respectively.
\end{lemma}

Now we take the unitary $U_D$ and apply it to the state $\ket{\bf 0}\ket{0}_{N^2}$ (where $\ket{0}_{N^2}$ denotes the first computational basis of a $N^2$-dimensional Hilbert space), according to Definition \ref{def: blockencode}, we have:
\begin{align}
    U_D \ket{\bf 0}\ket{0}_{N^2} = \frac{1}{ \alpha||D||^p} \ket{\bf 0}  \sum_{i,j=1}^N \ket{i-1} \exp\left( -\frac{d_{ij}^2}{||\sigma||^2} \right) \ket{j}  + \sum_{ k \neq \bf 0} \ket{k}\ket{\rm Garbage_k}
\end{align}
By tracing out the third register, we obtain the density state:
\begin{align}
    \rho = \ket{\bf 0}\bra{\bf 0} \otimes \frac{1}{ \alpha^2||D||^{2p}} K^\dagger K + \sum_{ k \neq \bf 0} \ket{k} \bra{k} \otimes \ket{\rm Garbage_k} \bra{\rm Garbage_k}
\end{align}
The above density state is again a block-encoding of $\varpropto  K^\dagger K$, and can be block-encoded via Lemma \ref{lemma: improveddme}.

\section{Quantum algorithm for obtaining geodesic distances}
\label{sec: detailgeodesicdistance}
We recall that the geodesic distance between two data points $\xbf_i,\xbf_j$ is approximated as:
\begin{align}
    d_G(\xbf_i,\xbf_j) \approx \left[  \sum^N_{k=1} \lambda_k^{2t}(\psi_{ik} - \psi_{jk})^2 \right]^{1/2}
\end{align}
where $\ket{\psi_k}_i $ refers to the $i$-th component of the $k$-th eigenvector $\ket{\psi_k}$ of $K$ and $\lambda_k$ is the corresponding eigenvalue of $K$. For convenience, we set $t=1$, so that the geodesic is further simplified as:
\begin{align}
    d_G(\xbf_i,\xbf_j) \approx \left[  \sum^N_{k=1} \lambda_k^{2}(\psi_{ik} - \psi_{jk})^2 \right]^{1/2}
\end{align}
We remark that from the previous section, we have the block-encoding of $\varpropto K^\dagger K$, which is also $\varpropto \sum_{k=1}^N \lambda_k^2 \ket{\psi_k}\bra{\psi_k}$. To proceed, we mention the following QSVT recipe:
\begin{lemma}[Negative Power Exponent \cite{gilyen2019quantum}, \cite{chakraborty2018power}]
\label{lemma: negative}
    Given a block encoding of a positive matrix $\frac{\mathcal{M}}{\gamma}$ such that 
    $$ \frac{\Ibb}{\kappa_M} \leq \frac{\mathcal{M}}{\gamma}\leq \Ibb. $$
    then we can implement an $\epsilon$-approximated block encoding of $\mathcal{M}^{-c}/(2\kappa_M^c)$ in complexity $\mathcal{O}( \kappa_M T_M (1+c) \log^2(  \frac{\gamma \kappa_M^{1+c}}{\epsilon} ) )$ where $T_M$ is the complexity to obtain the block encoding of $\mathcal{M}$. 
\end{lemma}
\noindent
We point out the following property:
\begin{align}
    \ket{\psi_k}_i - \ket{\psi_k}_j = e_{ij}^T \ket{\psi_k}
\end{align}
where $e_{ij}$ is the vector (of dimension $N$) that has entry 1 at position $i$-th, -1 at position $j$-th, and 0 otherwise. Let $E_i$ be the matrix having $j$-th row being $e_{ij}$ (for $j \neq i$), and for $i=j$, the whole row is zero. Then we have:
\begin{align}
    E_i  \sum_{k=1}^N \lambda_k^2 \ket{\psi_k}\bra{\psi_k} E_i = \sum_{k=1}^N \lambda_k^2\begin{pmatrix}
         \ket{\psi_k}_i - \ket{\psi_k}_1 \\
          \ket{\psi_k}_i - \ket{\psi_k}_2 \\
          \cdots \\
           \ket{\psi_k}_i - \ket{\psi_k}_N
    \end{pmatrix} \begin{pmatrix}
         \ket{\psi_k}_i - \ket{\psi_k}_1,  \ket{\psi_k}_i - \ket{\psi_k}_2, ...,  \ket{\psi_k}_i - \ket{\psi_k}_N
    \end{pmatrix}
\end{align}
The geodesic distance between $\xbf_i$ and $\xbf_j$ is the $j$-th diagonal entry of the above matrix. More generally, we consider the matrix $\sum_{i=1}^N \ket{i-1}\bra{i-1} \otimes E_i$, $\sum_{i=1}^N \ket{i-1}\bra{i-1}\otimes \sum_{k=1}^N \frac{1}{\lambda_k} \ket{\psi_k}\bra{\psi_k}  $ and their product:
\begin{align}
    \big( \sum_{i=1}^N \ket{i-1}\bra{i-1} \otimes E_i\big) \big(\sum_{i=1}^N \ket{i-1}\bra{i-1}\otimes \sum_{k=1}^N \lambda_k^2 \ket{\psi_k}\bra{\psi_k}  \big) \big( \sum_{i=1}^N \ket{i-1}\bra{i-1} \otimes E_i\big) 
\end{align}
The $(i\cdot j)$-th diagonal entry of the above matrix is exactly the geodesic distance $d_G(\xbf_i,\xbf_j)$ between the point $\xbf_i$ and $\xbf_j$. 

From the block-encoding of $\varpropto \sum_{k=1}^N \lambda_k^2\ket{\psi_k}\bra{\psi_k}  $ (as obtained from above), to obtain the block-encoding of $E_i$, we need the following result from \cite{nghiem2025refined}:
\begin{lemma}
\label{lemma: blockencodingknownmatrix}
    Suppose that $A$ is a matrix of size $N \times N$ with condition number $\kappa_A$, and that we are provided with classical knowledge/description of entries of $A$. Then the $\epsilon$-approximated block-encoding of $\frac{A}{||A||_F}$ can be obtained with a quantum circuit of complexity $\mathcal{O}\left( \log N \log^2 \frac{\kappa_A}{\epsilon} \right)$ and  a classical preprocesing of complexity $\mathcal{O}(\log N)$ (where $||A||_F$ is the Frobenius norm of $A$).
\end{lemma}
The application of the above lemma is straightforward to obtain the block-encoding of $\frac{\sum_{i=1}^N \ket{i-1}\bra{i-1}\otimes E_i  }{||E||_F}$ where $||E||_F$ is the Frobenius norm of the numerator. Using Lemma \ref{lemma: product}, we can obtain the block-encoding of:
\begin{align}
    &\frac{\sum_{i=1}^N \ket{i-1}\bra{i-1}\otimes E_i  }{||E||_F} \cdot  \big(\sum_{i=1}^N \ket{i-1}\bra{i-1}\otimes \sum_{k=1}^N \frac{\lambda_k^2}{\alpha||D||^p}\ket{\psi_k}\bra{\psi_k}  \big) \cdot  \frac{\sum_{i=1}^N \ket{i-1}\bra{i-1}\otimes E_i  }{||E||_F} \\
    &= \sum_{i=1}^N \ket{i-1}\bra{i-1} \otimes \frac{(\lambda^2_k)}{\alpha||D||^p||E||_F^2 } E_i \ket{\lambda_k}\bra{\lambda_k}E_i
    \label{c6}
\end{align}
which contains the square of geodesic distance (up to a scaling of Frobenius norm $||E||_F^2$) $d_G^2$ on the diagonal. Because each row of matrix $E_i$ has two non-zero entries being 1 and -1, so the Frobenius norm $||E_i||_F$ is $\sqrt{2N}$. The Frobenius norm of $E$ is $\sqrt{ 2N^2}= \sqrt{2}N$. For a reason that would be clear later, we only want to keep the diagonal entry. To ``filter'' out those off-diagonal entries, we can use the following procedure:
\begin{lemma}
\label{lemma: filteroffdiagonal}
    Let $U$ be a unitary block-encoding of some matrix $M$ of size $n \times n$. Let $T_U$ be the circuit complexity of $U$, then the block-encoding of $\sum_{i=1}^n \frac{1}{n} M^2_{ii} \ket{i}\bra{i}$ can be obtained with a circuit of depth $\mathcal{O}\left(  T_U + \log n \right)$. 
\end{lemma}
\noindent
\textbf{Proof:} By applying $U$ to the state $\ket{\bf 0} \frac{1}{\sqrt{n}} \sum_{i=1}^n \ket{i-1}$, we obtain the following state:
\begin{align}
    &\frac{1}{\sqrt{n}} \ket{\bf 0}\sum_{i=1}^n \ket{i-1 }M^i + \sum_{k \neq \bf 0} \ket{k} \ket{\rm Garbage} \\
    &= \frac{1}{\sqrt{n}} \ket{\bf 0}\sum_{i=1}^n  \sum_{j=1}^N M_{ij} \ket{i-1}\ket{j-1} + \sum_{k \neq \bf 0} \ket{k} \ket{\rm Garbage} 
\end{align}
where $M^i$ is the $i$-th column of $M$ and $M_{ij}$ its the $j$-th entry. Now we append another ancilla initialized in $\ket{0}^{\otimes \log n}$, and use the CNOT gates to obtain the following state:
\begin{align}
    \frac{1}{\sqrt{n}} \ket{\bf 0}\sum_{i=1}^{n}  \sum_{j=1}^{n} M_{ij} \ket{i-1}\ket{j-1} + \sum_{k \neq \bf 0} \ket{k} \ket{\rm Garbage} 
\end{align}
where the $\ket{\rm Garbage}$ contains a slight abuse of notation. If we trace out the ancilla, we obtain the following density state:
\begin{align}
    \ket{\bf 0}\bra{\bf 0} \otimes \frac{1}{n} \sum_{i=1}^{n} (M_{ii})^2 \ket{i-1}\bra{i-1} + \rho_{\rm Garbage}
\end{align}
The above density state can be block-encoded via Lemma \ref{lemma: improveddme}, and in fact, the above density state is also the block-encoding of $  \frac{1}{n} \sum_{i=1}^n (M_{ii})^2 \ket{i-1}\bra{i-1}$, which contains the diagonal entries only.  $\blacksquare$ \\

The application of the lemma above to our case is straightforward. Denote the operator in Eqn.~\ref{c6} is $M$, and its unitary block-encoding is $U$. Then using the above lemma enables us to obtain the block-encoding of $ \frac{1}{N^2} \sum_{p=1}^{N^2} (M_{pp})^2 \ket{p-1}\bra{p-1} $. We note that the diagonal entries of the matrix $M$ are the geodesic distances (divided by the Frobenius norm $||E||_F$), i.e.,
\begin{align}
  \frac{1}{N^2} \sum_{p=1}^{N^2} (E_{pp})^2 \ket{p-1}\bra{p-1} =    \frac{1}{\alpha||D||^p||E||_F^4 N^2} \sum_{i,j=1}^N \ket{i-1}\bra{i-1}\otimes d^4_G(\xbf_i,\xbf_j) \ket{j-1}\bra{j-1}
  \label{d12}
\end{align}
where in the above, we have decompose $\ket{p} = \ket{i}\ket{j}$.\\

Let $U_{p_j}$ denotes the $N$-dimensional permutation unitary such that:
\begin{align}
    \ket{0} \leftrightarrow \ket{i-1}
\end{align}
and the remaining basis permuted arbitrarily. This unitary can be constructed via Lemma \ref{lemma: permutation} with a depth $\mathcal{O}\left( \log N\right)$. We then use Lemma \ref{lemma: tensorproduct} to obtain the unitary $U_{p_j} \otimes \Ibb_N$, and then Lemma \ref{lemma: product} to construct the block-encoding of:
\begin{align}
    &\left( U_{p_j} \otimes \Ibb_N \right) \cdot  \frac{1}{\alpha ||D||^p ||E||_F^2 N^2} \ket{i-1}\bra{i-1}\otimes d^2_G(\xbf_i,\xbf_j) \ket{j-1}\bra{j-1} \\ & = \frac{1}{\alpha ||D||^p ||E||_F^2 N^2}  \sum_{j=1}^N \ket{0}\bra{0}\otimes d^4_G(\xbf_i,\xbf_j) \ket{j-1}\bra{j-1} + \frac{1}{\alpha ||D||^p ||E||_F^2 N^2}  \sum_{i,j=2}^N  \ket{i-1}\bra{i-1} \otimes  d^4_G(\xbf_i,\xbf_j) \ket{j-1}\bra{j-1}
\end{align}
which is also the block-encoding of the operator:
\begin{align}
     \frac{1 }{\alpha||D||^p||E||_F^4 N^2}  \sum_{j=1}^N d_G^4(\xbf_i,\xbf_j) \ket{j-1}\bra{j-1}
\end{align}
which essentially contains $ \varpropto d^4_G(\xbf_i,\xbf_j)$ on the diagonal. In the following section, we will show how to make use of this operator for our purposes.

\section{Quantum algorithm for estimating intrinsic dimension}
\label{sec: detailestimatingintrinsicdistance}
To find the intrinsic dimension at the point $\xbf_i$, we remind the following steps from \textbf{Step 2} in the Appendix \ref{sec: pipelineclassicalalgorithm}:
\begin{itemize}
    \item For each point, say $\xbf_i$, choose a fixed number (typically small) of nearest points in $d_G$ (note that now the definition of nearest refer to the geodesic distance instead of the Euclidean distance as in the previous step). Let $N_i$ denotes the set of those nearest points of $\xbf_i$. 
    \item Define the center of the neighborhood  of $\xbf_i$ as:
    \begin{align}
        \widetilde{\xbf}_i = \frac{1}{|N_i|} \sum_{j, \xbf_j \in N_i } \xbf_j
    \end{align}
    \item Define the centered-coordinate neighborhood at $\xbf_i$ as:
    \begin{align}
        \widetilde{N}_i = \{ \xbf_j- \widetilde{\xbf}_i\}_{j, \xbf_j \in N_i}
    \end{align}
    \item Define the matrix $C_i$ to be the matrix of size $|N_i| \times m$, where the rows of $C_i$ is in correspondence with $\widetilde{N}_i$. 
    \item Perform singular decomposition on $C_i$ and obtain a series of singular values $\sigma_1 \geq \sigma_2 \geq ... \geq \sigma_{|N_i|}$. 
\end{itemize}

\subsection{Finding $N_i$ nearest points to $\xbf_i$}
\label{sec: findingninearestpoints}
Our first challenge is to find those $N_i$ nearest points to the point $\xbf_i$. Recall that we have the unitary block-encoding of the following operator:
\begin{align}
  \frac{1}{\alpha ||D||^p ||E||_F^4 N^2}   \sum_{j=1}^N d_G^4(\xbf_i,\xbf_j) \ket{j-1}\bra{j-1}
\end{align}
which essentially contains $ \varpropto d_G^2(\xbf_i,\xbf_j)$ on the diagonal. First, we use Lemma \ref{lemma: negative} with $c=1/4$ to transform the block-encoded operator:
\begin{align}
   \frac{1}{\alpha ||D||^p ||E||_F^4 N^2}  \sum_{j=1}^N d_G^4(\xbf_i,\xbf_j) \ket{j-1}\bra{j-1} \longrightarrow \sum_{j=1}^N \frac{d_G(\min) }{d_G(\xbf_i,\xbf_j)} \ket{j-1}\bra{j-1}
\end{align}
where $d_G(\min) =\min \{  d_G(\xbf_i,\xbf_j)\}_{i,j=1}^N$. As the next step, we point out the recent result of \cite{nghiem2025refined}:
\begin{lemma}
\label{lemma: largesteigenvalues}
    Let $A$ be a Hermitian matrix of size $N \times N$ with a block-encoding unitary $U_A$ (of complexity $T_A$). Denote $\{ \lambda_i, \ket{\lambda_i}\}_{i=1}^N$ as its eigenvalues and corresponding eigenvectors. Assume that the order of eigenvalues obey $\lambda_1 > \lambda_2 > ... > \lambda_N$.  Then the value of $r$ highest eigenvalues $\lambda_1,\lambda_2, ..., \lambda_r$ can be estimated, sequentially, up to additive accuracy, $\epsilon$ in complexity:
    \begin{align}
     \mathcal{O}\left(  T_A\frac{1}{\Delta_1 \epsilon}  \log \left(\frac{N}{\epsilon}\right) \log \frac{1}{\epsilon}  \right), \mathcal{O}\left(  T_A\frac{1}{\Delta_2^2 \epsilon}  \log^2 \left(\frac{N}{\epsilon}\right) \log^2 \frac{1}{\epsilon}  \right), ...    \mathcal{O}\left(  T_A\frac{1}{\Delta_r^r \epsilon}  \log^r \left(\frac{N}{\epsilon}\right) \log^r \frac{1}{\epsilon}  \right)
    \end{align}
   respectively, where $\Delta_j \equiv |\lambda_{j+1} - \lambda_{j}| $ (for $j=1,2,...,r$) is the gap between largest eigenvalues. The eigenvector $\ket{\lambda_1},\ket{\lambda_2}, ..., \ket{\lambda_r}$ can be obtained in complexity 
    \begin{align}
       \mathcal{O}\left(  T_A\frac{1}{\Delta_1}  \log \left(\frac{N}{\epsilon}\right) \log \frac{1}{\epsilon}  \right), \mathcal{O}\left(  T_A\frac{1}{\Delta_2^2 }  \log^2 \left(\frac{N}{\epsilon}\right) \log^2 \frac{1}{\epsilon}  \right), ...    \mathcal{O}\left(  T_A\frac{1}{\Delta_r^r}  \log^r \left(\frac{N}{\epsilon}\right) \log^r \frac{1}{\epsilon}  \right)
    \end{align}
\end{lemma}
We refer the interested readers to the Appendix \ref{sec: reviewPCA} for a more detailed description of the quantum algorithm behind the lemma above. We point out that as the operator
\begin{align}
    \sum_{j=1}^N \frac{d_G(\min) }{d_G(\xbf_i,\xbf_j)} \ket{j-1}\bra{j-1}
\end{align}
is diagonal, its eigenvalues are $\{\frac{d_{\min} }{d_G(\xbf_i,\xbf_j)}  \}_{j=1}^N $ and its eigenvectors are the computational basis state. The maximum eigenvalues of the above operator corresponds to those minimum geodesic distances $d_G(\xbf_i,\xbf_j)$. The application of the above lemmas to our procedure is straightforward, as we can use Lemma \ref{lemma: largesteigenvalues} to find the top, say, $|N_i|$ eigenvalues of the above operator. Their eigenvectors are ideally those computational basis state $\{\ket{i}\}$ corresponding to these eigenvalues, and can also be revealed via Lemma \ref{lemma: largesteigenvalues}. However, there is a subtlety. The output of Lemma \ref{lemma: largesteigenvalues} is the approximation to the largest eigenvectors. For example, suppose that we obtain some state $\ket{\Tilde{i}}$ which is not the computational basis state, but rather the $\epsilon$-approximation of the ideal state $\ket{i}$. In order to obtain the knowledge of the underlying index $i$, we can perform measurement in the computational basis. As $|| \ket{\Tilde{i}} - \ket{i}|| \leq \epsilon $, the probability of measuring $\ket{i}$ is $\geq 1- \epsilon$. By performing the measurement a few times, we can obtain the real index $i$. All in all, we obtain the knowledge of $|N_i|$ smallest geodesic distances $\{ d_G(\xbf_i,\xbf_j)\}$, and indexes of those points, encoded in the computational basis state. 

\subsection{Obtaining the block-encoding of $\varpropto C_i^\dagger C_i$}
\label{sec: obtainblockencodingofcici}
Our first challenge is to somehow, from the classical knowledge of those points $\xbf_j$ obtained above, construct the centroid:
\begin{align}
        \widetilde{\xbf}_i = \frac{1}{|N_i|} \sum_{j, \xbf_j \in N_i } \xbf_j
    \end{align}
To proceed, we use Lemma \ref{lemma: stateprepration} and the classical knowledge of those points $\xbf_j \in N_i$ to prepare the following state:
\begin{align}
    \ket{\widetilde{\xbf}_i } = \frac{1}{ \sqrt{\sum_{j, \xbf_j \in N_i } ||\xbf_j||^2}}\sum_{j, \xbf_j \in N_i } \ket{j-1}\xbf_j
\end{align}
Denote this unitary by $U_{  \widetilde{\xbf}_i } $. It can be seen that the first column of this unitary is $ \sum_{j, \xbf_j \in N_i } \ket{j}\xbf_i$. We consider the Hadamard gates $H^{\otimes \log |N_i|}$, which is trivial to prepare. Then we use Lemma \ref{lemma: tensorproduct} to construct the block-encoding of $H^{\otimes \log |N_i|} \otimes \Ibb_m $ (where remind that $m$ is the dimension of the original space $X$). We then use Lemma \ref{lemma: product} to construct the block-encoding of:
\begin{align}
    \left(H^{\otimes \log |N_i|} \otimes \Ibb_m  \right) U_{\widetilde{\xbf}_i }
\end{align}
The first column of this operator is:
\begin{align}
      \left(H^{\otimes \log |N_i|} \otimes \Ibb_m  \right) \frac{1}{ \sqrt{\sum_{j, \xbf_j \in N_i } ||\xbf_j||^2}} \sum_{j, \xbf_j \in N_i  } \ket{j-1}\xbf_j &= \frac{1}{\sqrt{|N_i| } \sqrt{\sum_{j, \xbf_j \in N_i } ||\xbf_j||^2}}\ket{\bf 0}  \sum_{j, \xbf_j \in N_i } \xbf_j + \sum_{k \neq \bf 0} \ket{k} \ket{\rm Garbage}
\end{align}
where $\sum_{k \neq \bf 0} \ket{k} \ket{\rm Garbage} $ denotes the irrelevant part, which can be safely ignored. We only pay attention to the first $m$ entries of the first column, which is:
\begin{align}
    \frac{1}{\sqrt{|N_i|} \sqrt{\sum_{j, \xbf_j \in N_i } ||\xbf_j||^2}} \sum_{j, \xbf_j \in N_i } \xbf_j
\end{align}
Next, we consider the unitary $ H^{\otimes \log |N_i|}$, and use Lemma \ref{lemma: tensorproduct} to construct the block-encoding of:
\begin{align}
    H^{\otimes \log |N_i|} \otimes  \left(H^{\otimes \log |N_i|} \otimes \Ibb_m  \right) U_{\widetilde{\xbf}_i }
\end{align}
The first $|N_i| \times m$ entries of the first column of the above operator is:
\begin{align}
   \frac{1}{\sqrt{|N_i|}}  \sum_{j=1}^{|N_i|}\ket{j-1}  \frac{1}{\sqrt{|N_i|} \sqrt{\sum_{j, \xbf_j \in N_i } ||\xbf_j||^2}} \widetilde{\xbf}_i  =  \frac{1}{ \sqrt{\sum_{j, \xbf_j \in N_i } ||\xbf_j||^2}} \widetilde{\xbf}_i 
\end{align}
Recall that we have that the unitary $U_{\Tilde{\xbf}_i}$ contains the following state in the first column:
\begin{align}
     \ket{\widetilde{\xbf}_i } = \frac{1}{ \sqrt{\sum_{j, \xbf_j \in N_i } ||\xbf_j||^2}}\sum_{j, \xbf_j \in N_i } \ket{j-1}\xbf_j
\end{align}
Next, we use the unitary $U_{\widetilde{\xbf}_j}$ and the block-encoding of $  H^{\otimes \log |N_i|} \otimes  \left(H^{\otimes \log |N_i|} \otimes \Ibb_m  \right) U_{\widetilde{\xbf}_i }$ with Lemma \ref{lemma: sumencoding} to construct the block-encoding of their subtraction:
\begin{align}
    \frac{1}{2}\left(U_{\widetilde{\xbf}_j} -  \left( H^{\otimes \log |N_i|} \otimes  \left(H^{\otimes \log |N_i|} \otimes \Ibb_m  \right) U_{\widetilde{\xbf}_i }\right)_{|N_i| m \times |N_i| m} \right) 
    \label{eqn: e15}
\end{align}
where $ (.)_{|N_i| m \times |N_i| m} $ refers to the top-left corner matrix of size $|N_i| m \times |N_i| m $, i.e., the block-encoded matrix. The above block-encoded operator has the first column to be:
\begin{align}
    \frac{1}{2}  \frac{1}{ \sqrt{\sum_{j, \xbf_j \in N_i } ||\xbf_j||^2} } \left(  \sum_{j=1}^{|N_i|}\ket{j-1}    \left(\xbf_j-  \widetilde{\xbf}_i \right)\right)
\end{align}
We recall from earlier that we need to obtain the matrix $C_i$ where the rows of $C_i$ is corresponding to $  \widetilde{N}_i = \{ \xbf_j- \widetilde{\xbf}_i\}_{j, \xbf_j \in N_i}$. Our goal now is to build the block-encoding of $ C_i$, from the block-encoding of the above operator. Taking the above block-encoding and apply it to the state $\ket{\bf 0} \ket{0}_{m |N_i|}$ where $\ket{0}_{m |N_i}$ refers to the first computational basis state of the $(m |N_i|)$-dimensional Hilbert space and $\ket{\bf 0}$ refers to the ancilla qubits required for block-encoding purpose. According to Definition \ref{def: blockencode}, we obtain the following state:
\begin{align}
    \ket{\bf 0} \frac{1}{2}  \frac{1}{ \sqrt{\sum_{j, \xbf_j \in N_i } ||\xbf_j||^2}} \left(  \sum_{j=1}^{|N_i|} \ket{j-1}    \left(\xbf_j-  \widetilde{\xbf}_i \right)\right) + \sum_{k \neq \bf 0} \ket{k}\ket{\rm Garbage_k}
    \label{eqn: d22}
\end{align}
We consider the decomposition $  \left(\xbf_j-  \widetilde{\xbf}_i \right) = \sum_{k=1}^m  \left(\xbf_j-  \widetilde{\xbf}_i \right)_k  \ket{k-1}  $, and we consider the last two qubits register: 
\begin{align}
    \ket{j-1}    \left(\xbf_j-  \widetilde{\xbf}_i \right) = \ket{j-1} \sum_{k=1}^m  \left(\xbf_j-  \widetilde{\xbf}_i \right)_k  \ket{k-1} 
\end{align}
If we use the SWAP gates to swap these two register, we obtain the following state:
\begin{align}
     \ket{k-1}\sum_{k=1}^m  \left(\xbf_j-  \widetilde{\xbf}_i \right)_k  \ket{j-1}
\end{align}
Therefore, the state in Eqn.~\ref{eqn: d22} becomes:
\begin{align}
     \ket{\bf 0} \frac{1}{2}  \frac{1}{ \sqrt{\sum_{j, \xbf_j \in N_i } ||\xbf_j||^2}} \left(  \sum_{j=1}^{|N_i|} \sum_{k=1}^m    \ket{k-1}  \left(\xbf_j-  \widetilde{\xbf}_i \right)_k \ket{j-1} \right) + \sum_{k \neq \bf 0} \ket{k}\ket{\rm Garbage_k}
     \label{eqn: e20}
\end{align}
We point out the following crucial property:
\begin{align}
    \sum_{j=1}^{|N_i|} \sum_{k=1}^m    \ket{k-1}  \left(\xbf_j-  \widetilde{\xbf}_i \right)_k \ket{j-1} = \sum_{k=1}^m C_i^{k} \ket{k-1}
\end{align}
where we remind that $C_i$ is the matrix having $\{ \xbf_j - \widetilde{\xbf}_i \}$ as columns, and $C_i^k$ denotes the $k$-th column of $C_i$. Thus, tracing out the second register, we obtain the density state:
\begin{align}
    \frac{1}{4}  \frac{1}{\sum_{j, \xbf_j \in N_i } ||\xbf_j||^2}  \ket{\bf 0}\bra{\bf 0} \otimes \left( C_i^\dagger C_i\right) +  (...)
\end{align}
where $(...)$ denotes the irrelevant part. The above operator can be block-encoded via Lemma \ref{lemma: improveddme}, and at the same time, by Definition \ref{def: blockencode}, the above operator is a block-encoding of:
\begin{align}
     \frac{1}{4}  \frac{1}{\sum_{j, \xbf_j \in N_i } ||\xbf_j||^2}  \otimes \left( C_i^\dagger C_i\right) 
\end{align}
We point out that the factor $ \sum_{j, \xbf_j \in N_i } ||\xbf_j||^2$ can be removed via Lemma \ref{lemma: amp_amp}. Thus, we have obtained the block-encoding of $\frac{1}{2} C_i^\dagger C_i$. Our next goal is to estimate the local intrinsic dimension from the spectrum of this operator. 

\subsection{Estimating the (local) intrinsic dimension}
\label{sec: estimatinglocalintrinsicdimension}
We recall the last part of \textbf{Step 2} in the previous appendix \ref{sec: pipelineclassicalalgorithm}, which accounts for the (local) intrinsic dimension estimation:
\begin{itemize}
\item Define the matrix $C_i$ to be the matrix of size $|N_i| \times m$, where the rows of $C_i$ is in correspondence with $\widetilde{N}_i$. 
    \item Perform singular decomposition on $C_i$ and obtain a series of singular values $\sigma_1 \geq \sigma_2 \geq ... \geq \sigma_{|N_i|}$. 
    \item Estimate the local dimension $d_i$ in the neighborhood of $\xbf_i$ by finding the integer $k$ so that:
    \begin{align}
        d_i = \arg \min_k \left\{ k \ \Bigg|\  \frac{\sum_{\alpha=1}^k \lambda_\alpha }{\sum_{\alpha=1}^{|N_i|} \lambda_\alpha} = \frac{\sum_{\alpha=1}^k \sigma_\alpha^2 }{\sum_{\alpha=1}^{|N_i|} \sigma_\alpha^2} \geq \tau  \right\}
    \end{align}
    where $\tau \in (0.9,0.99)$ is the threshold. In other words, we require that the subspace spanned by $\{\vec{\mathbb{U}}_\alpha\}^{d_i}_{\alpha=1}$ to explain at least a fraction $\tau$ of the point-cloud's total variance.
\end{itemize}
To estimate $\sum_{\alpha=1}^k \sigma_\alpha^2 $, we can choose a value of $k$ and then use Lemma \ref{lemma: largesteigenvalues} to find the $k$ largest eigenvalues of $ \frac{1}{2}C_i^\dagger C_i$. The summation can then be done classically to obtain the estimate of the desired sum. To estimate the term $\sum_{\alpha=1}^{|N_i|} \sigma_\alpha^2 $, a naive approach would be choosing $k = |N_i|$ and use Lemma \ref{lemma: largesteigenvalues} to find all eigenvalues of $C_i^\dagger C_i$. While this approach is straightforward, we point out that in the Lemma \ref{lemma: largesteigenvalues}, the complexity is exponentially in the number of $k$. If we choose $k = |N_i|$, then the complexity will be exponential in $|N_i|$. For $|N_i| = \mathcal{O}(1)$, this cost is $\mathcal{O}(1)$, which is still modest, albeit the overhead/constant-factor is not small. However, we note that this step can be alternatively done, which is more efficient. The procedure is based on the following result of \cite{rall2020quantum}:
\begin{lemma}
\label{lemma: traceArho}
    Let $U_A$ be the unitary block-encoding of $A$, which is some Hermitian operator with $||A||_{o}\leq 1$ (where $||.||_{o}$ refers to operator norm) and $\rho$ be a density matrix of the same size as $A$. Let $U_\rho$ be the unitary preparing the state $\ket{\phi} $ s.t $\rho = \Tr_{H} \ket{\phi}\bra{\phi} $ ($\Tr_H$ refers to tracing out of some subsystem $H$). Let $T_A$ denotes the circuit complexity of $U_A$ and $T_\rho$ denotes the circuit complexity of $U_\rho$. Then there is a quantum procedure, using a circuit of depth $\mathcal{O}\left( \frac{1}{\epsilon} (T_A + T_\rho) \right)$ that returns an estimation of $ \Tr (A \rho)$ with an additive accuracy $\epsilon$. 
\end{lemma}
To apply the above lemma so as to estimate $ \sum_{\alpha=1}^{|N_i|} \sigma_\alpha^2$, we need to obtain the unitary block-encoding of $\sum_{i=1}^{|N_i|} \frac{1}{|N_i|}\ket{i}\bra{i}$. This can be done as follows. First, we take the Hadamard gates $H^{\otimes \log |N_i|}$ and apply it to $\ket{0}^{\otimes \log |N_i|}$, then we obtain the state $\frac{1}{\sqrt{|N_i|}}\sum_{i=1}^{|N_i|} \ket{i}$. Then we append the ancilla initialized in $\ket{0}^{\otimes \log |N_i|}$, and use CNOT gates to transform:
\begin{align}
    \frac{1}{\sqrt{|N_i|}}\sum_{i=1}^{|N_i|} \ket{i} \ket{0}^{\otimes \log |N_i|} \longrightarrow \frac{1}{\sqrt{|N_i|}}\sum_{i=1}^{|N_i|} \ket{i} \ket{i}
\end{align}
Tracing out either register yields $\sum_{i=1}^{|N_i|} \frac{1}{|N_i|}\ket{i}\bra{i} $. The unitary that prepares the above state consists of $\log |N_i|$ gates (applied in parallel) and a layer of $\log |N_i|$ CNOT gates. Thus, the circuit depth of this unitary is $\mathcal{O}(\log |N_i|)$. Therefore, the application of Lemma \ref{lemma: traceArho} to estimate the desired summation will incur a total complexity $\mathcal{O}(\log |N_i|)$ (we temporarily ignore the other dependence), which is significantly improved compared to the naive approach mentioned earlier. 

Last, to find the local dimension $d_i$, we need to find the value of $k$ at which the ratio $\frac{\sum_{\alpha=1}^k \sigma_\alpha^2 }{\sum_{\alpha=1}^{|N_i|} \sigma_\alpha^2} \geq \tau$. Our proposed strategy is that, we first pick a value of $k$, then find the $k$ largest eigenvalues of $\frac{1}{2}C_i^\dagger C_i$, and also the summation $ \sum_{\alpha=1}^{|N_i|} \sigma_\alpha^2$. Then we sequentially test, for $p=1,2,..,k$, the ratio:
\begin{align}
    \frac{\sum_{\alpha=1}^p \sigma_\alpha^2 }{\sum_{\alpha=1}^{|N_i|} \sigma_\alpha^2}
\end{align}
and find the value of $p$ at which the above ratio reaches $\tau \in (0.9,0.99)$, which gives the value of local dimension $d_i$.

\section{Quantum $\&$ classical procedure for estimating volume of geodesic balls}
\label{sec: quantumalgorithmestimatingdensity}
Recall that the so-called sampling density $\rho(\xbf_i)$ at the data point $\xbf_i$ is estimated as follows:
\begin{align}
    \rho(\xbf_i) \approx \sum_{j, \xbf_j \in N_i}  \exp\left( - \left( \frac{d_G(\xbf_i,\xbf_j)}{h}\right)^2 \right)
\end{align}
where $h$ is the scale parameter which controls the locality and $d_G$ is the geodesic distance.  In the previous section, it was shown that for a given point $\xbf_i$, we can leverage quantum PCA algorithm to find those $N_i$ nearest point to $\xbf_i$, alongside the estimation of the distances $\{d_G(\xbf_i,\xbf_j)  \}_{j, \xbf_j \in N_i} $. Therefore, the value of density above can be estimated using classical computer. The time complexity for this estimation would be $\mathcal{O}( |N_i|)$. As we pointed out from the main text, the value of $|N_i|$ is typically (and in fact, can be fixed) to be $\mathcal{O}(1)$. Thus, the classical complexity is of negligible cost. 

For a given point $\xbf_i$, the geodesic ball of radii $r$ around $\xbf_i$ as follows:
    \begin{align}
        B_r(\xbf_i) = \{ \xbf_j \in X | d_G(\xbf_i,\xbf_j) \leq r \}
    \end{align}
    The volume of the geodesic ball as follows:
    \begin{align}
        \rm Vol \left(  B_r(\xbf_i)\right) = \sum_{\xbf_j \in B_r(\xbf_i)} \frac{1}{\rho(\xbf_j)}
    \end{align}
The normalized volume of geodesic ball is estimated as:
    \begin{align}
        \rm Vol_{nor} \left(  B_r(\xbf_i)\right) &= \frac{\rm Vol \left(  B_r(\xbf_i)\right) }{ w_d r^d} \\
       \text{where \ }  w_d &= \frac{\pi^{d/2}}{\Gamma (\frac{d}{2} + 1 )}
    \end{align}
where we remind that the value of $d$ above is the local intrinsic dimension $d_i$ that we found in the previous step. Provided that the geodesic distances $\{ d_G(\xbf_i,\xbf_j) \}_{j, \xbf_j \in N_i} $ and the sampling density are known, then for a known value of radii $r$, the value of the normalized volume of the geodesic ball around $\xbf_i$ can be classically computed at $\mathcal{O}(1)$ cost. 

\section{Fitting quadratic curve to find the curvature at $\xbf_i$}
\label{sec: fitcurvature}
We recall the \textbf{Step 5} of the classical algorithm in Appendix \ref{sec: pipelineclassicalalgorithm}. First, choose a range of radii $r_1,r_2,..,r_M$ and find their corresponding volume of the geodesic ball $ \rm Vol_{nor} \left(  B_{r_1}(\xbf_i)\right),  \rm Vol_{nor} \left(  B_{r_2}(\xbf_i)\right), ...,  \rm Vol_{nor} \left(  B_{r_M}(\xbf_i)\right)   $. From these data, we perform the quadratic fit:
\begin{align}
     \rm Vol_{nor} \left(  B_{r}(\xbf_i)\right) \  \rm versus \  r^2
\end{align}
with the fit function being:
\begin{align}
     \rm Vol_{nor} \left(  B_r(\xbf_i)\right)  = 1 + A r^2
\end{align}
Our strategy for this fitting step is as follows. Consider the set $\{ \xbf_j\}_{j, \xbf_j \in N_i}  $ of $N_i$ nearest points to $\xbf_i$, we sort the geodesic distance from lowest to highest, via the classical algorithm. Then we treat these sorted geodesic distances as the radii $r_1,r_2,..., r_{|N_i|}$. As a result, the first geodesic ball $B_{r_1}(\xbf_i)$ will contains 2 points, including $\xbf_i$ and the closet point to it. The second geodesic ball $B_{r_2}(\xbf_i)$ will contain 3 points, and so on. The last geodesic ball $B_{r_{|N_i|}}(\xbf_i)$ will have all $|N_i|$ nearest points around $\xbf_i$. For each radii, the value of normalized volume can be computed classically, as we know the value of sampling density of all points inside it. 

To perform the quadratic fit, we aim to minimize the following cost function with $A$ being the parameter of interest:
\begin{align}
    C = \sum_{j=1}^{|N_i|} || 1+ A r_j^2 -  \rm Vol_{nor} \left(  B_{r_j}(\xbf_i)\right) ||^2
\end{align}
To find the value of $A$, we take the derivative of $C$ with respect to $A$ and set it equal to 0, e.g.:
\begin{align}
    \frac{\partial C}{\partial A} = \sum_{j=1}^{|N_i|} 2\left( 1+ A r_j^2 -  \rm Vol_{nor} \left(  B_{r_j}(\xbf_i)\right) \right)= 0
\end{align}
Then the value of $A$ can be found as:
\begin{align}
    A = \frac{\sum_{j=1}^{|N_i|}\rm Vol_{nor} \left( B_{r_j}(\xbf_i)\right) }{|N_i| + \sum_{j=1}^{|N_i|} d_G(\xbf_i,\xbf_j)^2}
\end{align}
The value of $A$ above can be expressed as:
\begin{align}
    A = \frac{ \sum_{j=1}^{|N_i|}\rm Vol_{nor} \left( B_{r_j}(\xbf_i) \right)/|N_i| }{1+ \sum_{j=1}^{|N_i|} d_G^2(\xbf_i,\xbf_j)/|N_i|}
\end{align}
Since the value of geodesic ball volumes and the geodesic distances are classically known, we can use classical procedure to compute the summations $  \sum_{j=1}^{|N_i|} \rm Vol_{nor}  B_{r_j}(\xbf_i) /|N_i|$ and $\sum_{j=1}^{|N_i|} d_G^2(\xbf_i,\xbf_j)/|N_i|$, respectively. This classical procedure will take $\mathcal{O}( |N_i|)$ times. At the same time, from the classical knowledge, these summations can be estimated using a quantum circuit of depth $\mathcal{O}(\log |N_i|) $ as follows. First, we use Lemma \ref{lemma: stateprepration} to prepare the states:
\begin{align}
    \frac{1}{||B||}  \sum_{j=1}^{|N_i|} \rm Vol_{nor} \left( B_{r_j}(\xbf_i)\right) \ket{j}, \frac{1}{||D_G||} \sum_{j=1}^{|N_i|} d_G^2(\xbf_i,\xbf_j) \ket{j}
\end{align}
where $||B|| \equiv \sqrt{ \sum_{j=1}^{|N_i|}B_{r_j}^2(\xbf_i) }$, $||D_G|| \equiv \sqrt{ \sum_{j=1}^{|N_i|} d_G^4(\xbf_i,\xbf_j) }$. This preparation use a circuit of depth $\mathcal{O}(\log |N_i|)$. Then, we prepare the state $\frac{1}{\sqrt{|N_i|}}\sum_{j=1}^{|N_i|} \ket{j}$, which can be done by a circuit $H^{\otimes \log |N_i|}$, which is depth 1. Next, we use Hadamard test/SWAP test to evaluate the inner products:
\begin{align}
  \left( \frac{1}{\sqrt{|N_i|}} \sum_{j=1}^{|N_i|} \bra{j} \right)   \left( \frac{1}{||B||}  \sum_{j=1}^{|N_i|}\rm Vol_{nor} \left( B_{r_j}(\xbf_i)\right) \ket{j}\right) = \frac{ 1}{\sqrt{|N_i| ||B||}} \sum_{j=1}^{|N_i|} \rm Vol_{nor} \left( B_{r_j}(\xbf_i) \right) \\
  \left(  \frac{1}{\sqrt{|N_i|}}\sum_{j=1}^{|N_i|} \bra{j} \right) \left(  \frac{1}{||D_G||} \sum_{j=1}^{|N_i|} d_G^2(\xbf_i,\xbf_j) \ket{j}\right) = \frac{ 1}{\sqrt{|N_i| ||D_G||}} \sum_{j=1}^{|N_i|}d_G^2(\xbf_i,\xbf_j)
\end{align}
The desired summation can be inferred by dividing the above values by $\sqrt{|N_i|}$ then multiply it with $||B||, ||D_G||$, respectively. 

\section{Complexity analysis}
\label{sec: complexityanalysis}
Here, we explicitly analyze the complexity of the procedure described above. A summary of this algorithm can be found in the diagram \ref{fig: Diagram}. The analysis we provide below reflects exactly the order of this diagram. 
\begin{enumerate}
    \item \textbf{Obtain the block-encoding of $K^\dagger K$ from the pairwise distance $\{ d(\xbf_i,\xbf_j)\}_{i,j=1}^N$. } In this part, we first need to use Lemma \ref{lemma: stateprepration} to prepare a $N$-dimensional state $\ket{\phi}$, which incurs complexity $\mathcal{O}\left( \log N\right)$. We then use Lemma \ref{lemma: permutation} and Lemma \ref{lemma: pairwisesquaregeneralized} to obtain the block-encoding of the vectors in Eqn.~\ref{eqn: power}. Thus the complexity is $\mathcal{O}\left( p \log N\right)$. Next, we use Lemma  \ref{lemma: sumencoding} to construct the block-encoding of their summation, incurring a further complexity $\mathcal{O}(p)$, thus the total complexity for obtaining the block-encoding of:
    \begin{align}
    \frac{1}{ \alpha||D||^p}  \sum_{i,j=1}^N \ket{i-1} \exp\left( -\frac{d_{ij}^2}{\sigma^2} \right) \ket{j} 
\end{align}
is $\mathcal{O}\left( p^2 \log N \right)$. Last, Lemma \ref{lemma: improveddme} is used to obtain the block-encoding of $\frac{1}{\alpha ||D||^p} K^\dagger K$, resulting in total complexity $\mathcal{O}\left( p^2 \log N \right)$. For $p = \mathcal{O}\left( \log 1/\epsilon\right)$, the polynomial $P_p(x)$ is $\epsilon$-approximated to the $\exp(-x^2)$, so the block-encoding of $\frac{1}{\alpha ||D||^p} K^\dagger K $ is $\epsilon$-approximated, with complexity $ \mathcal{O}\left( \log^2\left(\frac{1}{\epsilon} \right) \log N \right)$.

\item \textbf{Obtain the block-encoding of $  \frac{1}{\alpha ||D||^{2p} ||E||_F^2 N^2}  \sum_{j=1}^N d_G^4(\xbf_i,\xbf_j) \ket{j-1}\bra{j-1}$}. First, twe use Lemma \ref{lemma: blockencodingknownmatrix} to obtain the block-encoding of the matrix $E = \frac{\sum_{i=1}^N \ket{i-1}\bra{i-1}\otimes E_i  }{||E||_F} $. This incurs a complexity $\mathcal{O}\left( \log (N) \log^2\left( \frac{\kappa_E}{\epsilon}\right)\right)$ where $\kappa_E$ is the condition number of $E$. We note that precisely, Lemma \ref{lemma: blockencodingknownmatrix} requires a classical pre-processing step of complexity $\mathcal{O}(\log N)$. However, we point out that, the matrices $\{ E_i \}$ have entries to be either 1 or 0, thus, the cost of pre-processing can be reduced to $\mathcal{O}(1)$. 

Next, we use Lemma \ref{lemma: product} to construct the block-encoding of $\varpropto E (K^\dagger K) E   $, and then Lemma \ref{lemma: filteroffdiagonal} to filter out those off-diagonal elements. The result is a block-encoding of an operator which contains the geodesic distance $\{d_G^4(\xbf_i, \xbf_j)\}$ (up to some scaling) on the diagonal. This results in total complexity 
\begin{align}
    \mathcal{O}\left( \log (N) \log^2\left( \frac{\kappa_E}{\epsilon}\right)+ \log^2\left(\frac{1}{\epsilon} \right) \log N  \right) = \mathcal{O}\left( \log (N) \log^2\left( \frac{\kappa_E}{\epsilon}\right)  \right)
\end{align}
We then use permutation unitary in Lemma \ref{lemma: permutation} and Lemma \ref{lemma: filteroffdiagonal} to obtain the block-encoding of:
\begin{align}
     \frac{1}{\alpha||D||^p||E||_F^4 N^2}  \sum_{j=1}^N d_G^4(\xbf_i,\xbf_j) \ket{j-1}\bra{j-1}
\end{align}
As Lemma \ref{lemma: permutation}, \ref{lemma: filteroffdiagonal} only incurs further complexity $\mathcal{O}(\log N)$, so the total complexity up to this step is:
\begin{align}
    \mathcal{O}\left( \log (N) \log^2\left( \frac{\kappa_E}{\epsilon}\right)  \right)
\end{align}
where we remind that $(\lambda^2_{\min}) $ is the minimum eigenvalue of the kernel matrix $K$ and $\kappa_E$ is the condition number of matrix $E$. 
\item \textbf{Finding the neighborhood of $\xbf_i$.} The neighborhood of $\xbf_i$ is defined as $|N_i|$ closest points to $\xbf_i$, in terms of geodesic distance. As mentioned, we first use Lemma \ref{lemma: negative} to obtain the block-encoding of:
\begin{align}
     \sum_{j=1}^N \frac{d_G({\min}) }{d_G(\xbf_i,\xbf_j)} \ket{j-1}\bra{j-1}
\end{align}
where $d_G(\min) =\min \{  d_G(\xbf_i,\xbf_j)\}_{i,j=1}^N$. The complexity of this application of Lemma \ref{lemma: negative} to obtain the above operator is:
\begin{align}
     \mathcal{O}\left( d^4_G(\min) \log^2\left(\frac{\kappa_E}{\epsilon} \right) \log (N)\log^2 \left( \frac{\alpha ||D||^p ||E||_F^4 N^2}{ \epsilon}\right) \right)
\end{align}
Next, we use Lemma \ref{lemma: largesteigenvalues} to find the $|N_i|$ largest eigenvalues of the above operator, which correspond to those $|N_i|$ smallest values of $d_G(\xbf_i,\xbf_j)$. As stated in Lemma \ref{lemma: largesteigenvalues}, by choosing the error tolerance to be $\epsilon$, the complexity of this step is:
\begin{align}
     \mathcal{O}\left(d^4_G(\min) \log^2\left(\frac{\kappa_E}{\epsilon} \right) \log (N)\log^2 \left( \frac{\alpha ||D||^p ||E||_F^4 N^2}{ \epsilon}\right) \log^{|N_i|}\left( \frac{N}{\epsilon} \right) \left(\frac{1}{\epsilon \Delta^{|N_i|}}\right) \log^{|N_i|}\frac{1}{\epsilon} \right)
\end{align}   
where for convenience, we set $\Delta $ to be the maximum separation between two largest eigenvalues of   $ \sum_{j=1}^N \frac{d_{\min} }{d_G(\xbf_i,\xbf_j)} \ket{j-1}\bra{j-1}$, among top $N_i$ largest eigenvalues of this operator. In other words, if we denote the set $\mathscr{D}_1\equiv  \{d_G(\xbf_i,\xbf_j) \}_{j=1}^N $, and $\min \mathscr{D}_1$, then in an iterative manner, we define:
\begin{align}
        \Delta_j = | \mathscr{D}_j{\min} -  ( \mathscr{D}_j\backslash \min\mathscr{D}_j )_{\min}  | \\
        \rm define \ \mathscr{D}_{j+1} \equiv ( \mathscr{D}_j \backslash \min\mathscr{D}_j )
\end{align}
The value of $\Delta$ is defined as $\Delta \equiv  \max \{ \Delta_1,\Delta_2,... \Delta_{|N_i|} \}$. We point out that, as the eigenvalues of the above diagonal operator is essentially $\sim (d_G(\xbf_i,\xbf_j)^{-1}$, the value of $\Delta$  can be alternatively, albeit more conveniently, set as $\Delta$ as $\Delta = \min \{  d_G(\xbf_i,\xbf_j) - d_G(\xbf_i,\xbf_q)\}_{i,j,q} $.

\item \textbf{Obtaining the block-encoding of $ \frac{1}{4}  \frac{1}{\sum_{j, \xbf_j \in N_i } ||\xbf_j||^2} \left( C_i^\dagger C_i\right) $. } In this step, we first need to use Lemma \ref{lemma: stateprepration} to obtain the unitary $U_{\widetilde{\xbf}_i}$, which has complexity $\mathcal{O}\left( \log |N_i|  \right)$. Next, we need to use Lemma \ref{lemma: sumencoding} (in Eqn.~\ref{eqn: e15}), which has complexity $\mathcal{O}\left( \log |N_i| \right)$ as the gates $H^{\otimes \log |N_i|}$ has depth 1. In the step of Eqn.~\ref{eqn: e20}, we need to SWAP to quantum systems of dimension $m$ and $|N_i|$, thus requiring to use $\mathcal{O}\left( \log \max( |N_i|, m) \right)$ SWAP gates. In reality, the value of $|N_i|$ is typically $\ll m$, so we take the maximum value to be $m$. Thus, the total complexity after this step is $\mathcal{O}(\log m|N_i|)$. The next step use Lemma \ref{lemma: improveddme} to obtain the block-encoding of $\varpropto C_i^\dagger C_i$, which is a matrix of dimension $m\times m$, thus incurring a total complexity $\mathcal{O}\left( \log (m|N_i|)\right)$.

\item \textbf{Estimating the local dimension $d_i$.} In this step, we need to perform principal component analysis on $\varpropto C_i^\dagger C_i$ to find the local dimension $d_i$. Thus, Lemma \ref{lemma: largesteigenvalues} can be applied. Because the local dimension is defined to be the minimum integer $k$ s.t. $\frac{\sum_{\alpha=1}^k \sigma_\alpha^2 }{\sum_{\alpha=1}^{|N_i|} \sigma_\alpha^2} \geq \tau $, we need to use Lemma \ref{lemma: largesteigenvalues} at least $d_i$ times to find the largest $d_i$ eigenvalues. The total complexity is then $\mathcal{O}\left( \log (m|N_i|) \frac{1}{\delta^{d_i} \epsilon } \log^{d_i} \left( \frac{m}{\epsilon} \right) \log \left(\frac{1}{\epsilon}\right)  \right)$ where $\delta$ now is defined to be the maximum gap between two consecutive largest eigenvalues of $C_i^\dagger C_i$, among $d_i$ largest ones. 

\item \textbf{Fit the quadratic curve to infer the curvature. } Before fitting the quadratic curve, we need to use classical computer to compute the sampling density $\rho(\xbf_i)$, which is efficient. The value of geodesic ball and its normalization by a unit ball volume can thus be computed accordingly. The final step is to compute the quantities $\sum_{j=1}^{|N_i|}\rm Vol_{nor} \left( B_{r_j}(\xbf_i) \right)/|N_i|  $ and $ \sum_{j=1}^{|N_i|} d_G^2(\xbf_i,\xbf_j)/|N_i|$, in which we have provided two solutions, that either we use classical summation method (complexity $\mathcal{O}(|N_i)$) or we use the state preparation plus Hadamard/SWAP test procedure ($\mathcal{O}\left( \frac{1}{\epsilon}\log |N_i|\right)$ for an estimation with precision $\epsilon$). The fitting parameter can be found as:
\begin{align}
    A = \frac{ \sum_{j=1}^{|N_i|}\rm Vol_{nor} \left( B_{r_j}(\xbf_i) \right)/|N_i| }{1+ \sum_{j=1}^{|N_i|} d_G^2(\xbf_i,\xbf_j)/|N_i|}
\end{align}
Finally, the curvature can be estimated as $S(\xbf_i) = -6(d_i+2) A$. To sum up, the total complexity from the beginning till the estimation of curvature is:
\begin{align}
    \begin{split}
            \mathcal{O}\Big(d^4_G(\min) \log^2\left(\frac{\kappa_E}{\epsilon} \right) \log (N)\log^2 \left( \frac{\alpha ||D||^p ||E||_F^4 N^2}{ \epsilon}\right) \log^{|N_i|}\left( \frac{N}{\epsilon} \right) \left(\frac{1}{\epsilon \Delta^{|N_i|}}\right) \log^{|N_i|}\frac{1}{\epsilon} \\ 
            + \log (m|N_i|) \frac{1}{\delta^{\lceil d_i\rceil} \epsilon } \log^{\lceil d_i \rceil} \left( \frac{m}{\epsilon} \right) \log \left(\frac{1}{\epsilon}\right) \Big)
        \end{split}
\end{align}   
\end{enumerate}

As can be seen from the complexity above, our algorithm's performance depends on a few factors, specifically $|N_i|$, $d_i$, $||D||$, $||E||_F$ $d_G(\min)$ and $\Delta$ (as we set error tolerance $\epsilon$ to be a fixed value). As mentioned, the value of $|N_i|$ can be chosen, as $N_i$ defines the local neighborhood around $\xbf_i$, thus it is safe to treat it as $\mathcal{O}(1)$. Roughly speaking, $|N_i|$ is the upper bound to the local dimension. Thus, if the data points exhibit low-dimensional dimension, then the value of $|N_i|$ can be chosen to be small. The value of $||D||$ is equal to $\sqrt{\sum_{i,j=1}^N d(\xbf_i,\xbf_j)^2} $. In the worst case, if each $d(\xbf_i,\xbf_j)$ has $\mathcal{O}(1)$ magnitude, then $||D|| =\mathcal{O}( \sqrt{N})$. The value of $\lambda_{\min}$ depends on the spectrum of the kernel $K$, or more precisely, the condition number. The entries of $K$ depends on the pairwise distances among data points, therefore, in general, the value of $\lambda_{\min}$ depends on the data set. We thus set it to be $\mathcal{O}(1)$. The value of $||E||_F$ was pointed out earlier to be $\mathcal{O}(\sqrt{N})$. The value of $\delta$, depends on the spectrum of $C_i^\dagger C_i$, which again relies on the data set, thus can be safely considered to be $\mathcal{O}(1)$. The value of $\Delta$ is $ \min \{  d_G(\xbf_i,\xbf_j) - d_G(\xbf_i,\xbf_q)\}_{i,j,q} $, and $d_G(\min) = \min \{ d_G(\xbf_i, \xbf_j) \}_{j=1}^N$, which depends on the geodesic distances, thus can be safely treated as $\mathcal{O}(1)$. Summing up everything, the complexity can be simplified as:
\begin{align}
    \mathcal{O}\left(  \frac{1}{\Delta^{|N_i|} \epsilon} \log^{|N_i|+3} \left( N\right) + \frac{1}{\delta^{d_i} \epsilon } \log\left(m |N_i|\right) \log^{\lceil d_i \rceil} m  \right)
\end{align}

\section{Diffusion map algorithm}
\label{sec: diffusionmap}
\subsection{Classical algorithm}
The pipeline of (classical) diffusion map algorithm first proposed in \cite{coifman2006diffusion} is as follows.
\begin{method}[Diffusion Map]
Let $X = \{ \xbf_1,\xbf_2, ..., \xbf_N \} \subseteq \Rbb^m$ be the dataset and $d(\xbf_i,\xbf_j)$ is the pairwise distance (a form of similarity measure) between $\xbf_i$ and $\xbf_j$. 
\end{method}
\noindent
\textbf{Step 1: Define kernel.} The kernel matrix $K$ is defined as:
\begin{align}
    K_{ij} = \exp\left( - \frac{d(\xbf_i,\xbf_j)^2}{\sigma^2}\right)
\end{align}

\noindent
\textbf{Step 2: Normalization.} Compute the local density $q(\xbf_i) = \sum_{j} K_{ij}$, then normalize the kernel:
\begin{align}
    \widetilde{K}_{ij} = \frac{K_{ij}}{q(\xbf_i) q(\xbf_j)}
\end{align}

\noindent
\textbf{Step 3: Constructing diffusion operator.} Define the row-stochastic matrix, which is also a Markov chain:
\begin{align}
  P_{ij} =   p(\xbf_i,\xbf_j) = \frac{\widetilde{K}_{ij}  }{ \sum_j \widetilde{K}_{ij} }
\end{align}

\noindent
\textbf{Step 4: Spectral decomposition.} Diagonalize the matrix $P$, obtaining the eigenvalues/eigenvectors $\{ \lambda_k, \ket{\phi_k}\}_{i=1}^N$. Without loss of generalization, assume them to be in transcending order $\lambda_1 \geq \lambda_2 \geq ... \geq \lambda_N$. \\

\noindent
\textbf{Step 5: Embedding via diffusion coordinates. } Let $\phi_k(\xbf_i) \equiv (\phi_k)_i$ denotes the $i$-th component of the vector $\phi_k$. Define the diffusion map at time $t$ as:
\begin{align}
    \psi_t( \xbf_i) = \begin{pmatrix}
        \lambda_1^t \phi_1(\xbf_i), \lambda_2^t \phi_2(\xbf_i), ..., \lambda_n^t \phi_n(\xbf_i)
    \end{pmatrix}^T
\end{align}
This is a $m$-dimensional vector, which is regarded as the embedding/low-dimensional projection of $\xbf_i$.

\subsection{Quantum algorithm}
\noindent
\textbf{Algorithm.} Previously, in section \ref{sec: blockencodingkernelmatrix}, we have shown how to obtain the block-encoding of a matrix, denote as $M_K$, having the following vector as the first column:
\begin{align}
    \frac{1}{ \alpha||D||^p}  \sum_{i,j=1}^N \ket{i-1} \exp\left( -\frac{d_{ij}^2}{\sigma^2} \right) \ket{j-1} = \frac{1}{ \alpha||D||^p}  \sum_{i,j=1}^N \ket{i-1} K_{ij} \ket{j-1} 
\end{align}
From which the block-encoding of $\varpropto K^\dagger K$ can be obtained via Lemma \ref{lemma: improveddme}. 

We define the matrix $E$ of size $N^2 \times N^2$ as follows (note that this matrix $E$ is different from the previous sections). Within the first $N$ rows of $E$, the $i$-th row of $E$ has nonzero entries being 1, and their column indexes are $i,i+1,i+2,...,i+N-1$. For the remaining rows beyond $N$, all the entries are zero. According to Lemma \ref{lemma: blockencodingknownmatrix}, the matrix $\frac{E}{||E||_F}$ can be efficiently block-encoded. Then we use Lemma \ref{lemma: product} to construct the block-encoding of $ \frac{E}{||E||_F} \cdot M_K$. The product of this matrix $E/||E||_F$ with $M_K$ is another matrix having the following vector in the first column (first $N$ entries):
\begin{align}
    \frac{1}{  \alpha||E||_F||D||^p} \sum_{i=1}^N  ( \sum_{j=1}^N K_{ij} ) \ket{i-1} = \frac{1}{ \alpha||E||_F||D||^p} \sum_{i=1}^N  q(\xbf_i) \ket{i-1}
\end{align}
If we take this block encoded operator and apply it to the state $\ket{\bf 0}\ket{0}_N$ (where $\ket{0}_N$ denotes the first computational basis state of the $N$ dimensional Hilbert space), according to Definition \ref{def: blockencode}, we obtain the following state:
\begin{align}
   \ket{\beta} = \ket{\bf 0}  \frac{1}{ \alpha ||E||_F||D||^p} \sum_{i=1}^N  q(\xbf_i) \ket{i-1} + \ket{\rm Garbage}
\end{align}
To proceed, we need the following result:
\begin{lemma}[Theorem 2 in \cite{rattew2023non}, \cite{guo2024nonlinear}]
\label{lemma: diagonal}
     Given an $\log(n)$-qubit quantum state specified by a state-preparation-unitary $U$, such that $\ket{\psi}_n=U\ket{0}_n=\sum^{n}_{k=1}\psi_k \ket{k-1}_n$ (with $\psi_k \in \mathbb{C}$), we can prepare an exact block-encoding $U_A$ of the diagonal matrix $A = {\rm diag}(\psi_1, ...,\psi_{n})$ with $\mathcal{O}(\log(n))$ circuit depth and a total of $\mathcal{O}(1)$ queries to a controlled-$U$ gate  with $\log(n)+3$ ancillary qubits.
\end{lemma}
Applying the above lemma to the unitary that generates the state $\ket{\beta}$, we can obtain the block-encoding of the diagonal operator:
\begin{align}
     \frac{1}{ \alpha ||E||_F||D||^p} \sum_{i=1}^N  q(\xbf_i) \ket{i-1} \bra{i-1}
\end{align}
Using Lemma \ref{lemma: tensorproduct}, we can construct the block-encoding of:
\begin{align}
    \left( \frac{1}{ \alpha ||E||_F||D||^p} \sum_{i=1}^N  q(\xbf_i) \ket{i-1} \bra{i-1}\right) \otimes \left(  \frac{1}{ \alpha ||E||_F||D||^p} \sum_{i=1}^N  q(\xbf_i) \ket{i-1} \bra{i-1}\right) 
\end{align}
which is:
\begin{align}
    \left(\frac{1}{ \alpha ||E||_F||D||^p} \right)^2 \sum_{i,j=1}^N q(\xbf_i) q(\xbf_j) \ket{i-1}\bra{j-1}
\end{align}
Then, we use Lemma \ref{lemma: negative} with $c=1$ to inverse the above operator, i.e., we obtain the block-encoding of:
\begin{align}
    \sum_{i,j=1}^N \frac{q_{\min}}{ q(\xbf_i) q(\xbf_j)} \ket{i-1}\bra{j-1}
\end{align}
where $q_{\min} = \min \{  q(\xbf_i) q(\xbf_j)  \}_{i,j=1}^N$. We then use Lemma \ref{lemma: product} to take this unitary block-encoding and the unitary block-encoding of $M_K$, to construct the block-encoding of:
\begin{align}
   \left( \sum_{i,j=1}^N \frac{q_{\min}}{ q(\xbf_i) q(\xbf_j)} \ket{i-1}\bra{j-1} \right) \cdot M_k
\end{align}
As the first column of $M_K$ is $\frac{1}{ \alpha||D||^p}  \sum_{i,j=1}^N \ket{i-1} K_{ij} \ket{j-1} $, the first column of the above operator, denoted as $N_k$, is:
\begin{align}
    \frac{1}{ \alpha||D||^p} \sum_{i,j} \ket{i-1}  \frac{q_{\min} K_{ij}}{q(\xbf_i) q(\xbf_j)}\ket{j-1} = \frac{q_{\min}}{ \alpha||D||^p} \sum_{i,j} \ket{i-1}  \widetilde{K}_{ij}\ket{j-1}
\end{align}
From this block-encoding, we can repeat the same procedure as the beginning of this section, to obtain the block-encoding of the diagonal matrix:
\begin{align}
   \sum_{i,j=1}^N \frac{ w_{\min} }{ w(\xbf_i) }\ket{i-1}\bra{i-1}
\end{align}
where $w(\xbf_i)$ is defined as $w(\xbf_i) = \sum_{j=1}^N \widetilde{K}_{ij} $ and $w_{\min} = \min \{  w(\xbf_i) \}_{i=1}^N$. We then first use Lemma \ref{lemma: tensorproduct} to obtain the block-encoding of the above operator tensor producted with $\Ibb_N$, then use Lemma \ref{lemma: product} again to obtain the block-encoding of:
\begin{align}
   \left(\sum_{i,j=1}^N \frac{ w_{\min} }{ w(\xbf_i) }\ket{i-1}\bra{i-1}\otimes \Ibb_N\right) \cdot N_k
\end{align}
which contains the following vector in the first $N^2$ entries in the first column:
\begin{align}
    \frac{w_{\min} q_{\min}}{ \alpha||D||^p} \sum_{i,j=1}^N \ket{i-1} \frac{\widetilde{K}_{ij}}{w(\xbf_i) }\ket{j-1}
\end{align}
Then, we can follow the same procedure from Lemma \ref{lemma: improveddme} to obtain the block-encoding of:
\begin{align}
   \left( \frac{w_{\min} q_{\min}}{ \alpha||D||^p}\right)^2 P^\dagger P
\end{align}
where we remind that:
\begin{align}
  P_{ij} = \frac{\widetilde{K}_{ij}  }{ \sum_{j=1}^N \widetilde{K}_{ij} } = \frac{\widetilde{K}_{ij}  }{ w(\xbf_i) } 
\end{align}
To proceed, we point out the following result: 
\begin{lemma}[Positive Power Exponent \cite{gilyen2019quantum},\cite{chakraborty2018power}]
\label{lemma: positive}
    Given a block encoding of a positive matrix $\mathcal{M}/\gamma$ such that 
    $$ \frac{\Ibb}{\kappa_M} \leq \frac{\mathcal{M}}{\gamma} \leq \Ibb. $$
   Let $c \in (0,1)$. Then we can implement an $\epsilon$-approximated block encoding of $(\mathcal{M}/\gamma)^c/2$ in time complexity $\mathcal{O}( \kappa_M T_M \log^2 (\frac{  \kappa_M}{\epsilon})  )$, where $T_M$ is the complexity to obtain the block encoding of $\mathcal{M}/\gamma$. 
\end{lemma}
We then use the above lemma with $c= \frac{t}{2}$ to obtain the transformation on the block-encoded operator:
\begin{align}
    \left( \frac{w_{\min} q_{\min}}{ \alpha||D||^p}\right)^2 P^\dagger P \longrightarrow \frac{w_{\min} q_{\min}}{ \alpha||D||^p} P^t
\end{align}
From the block-encoding of the above operator, we then use Lemma \ref{lemma: largesteigenvalues} to find its largest $n$ eigenvalues, which are 
\begin{align}
    \frac{w_{\min} q_{\min}}{ \alpha||D||^p} \lambda_1^t, \frac{w_{\min} q_{\min}}{ \alpha||D||^p}\lambda_2^t ,..., \frac{w_{\min} q_{\min}}{ \alpha||D||^p}\lambda_n^t
\end{align}
with the corresponding eigenvectors $\ket{\phi_1},\ket{\phi_2},...,\ket{\phi_n}$. The $i$-th entry of these vectors can be obtained by measuring these states in the computational basis, and estimate the probability of measuring $\ket{i}$. Additionally, if we wish to obtain an estimate of any $\lambda_k^t$ (for $k=1,2,...,n$) to an additive precision $\epsilon$, then we need to estimate $ \frac{w_{\min} q_{\min}}{ \alpha||D||^p} \lambda_k^t $ to an accuracy $ \frac{w_{\min} q_{\min}}{ \alpha||D||^p} \epsilon$. \\

\noindent
\textbf{Complexity.} From the previous appendix, we have that the complexity of obtaining the block-encoding of $ \frac{1}{ \alpha||D||^p}  \sum_{i,j=1}^N \ket{i-1} K_{ij} \ket{j-1} $ is $\mathcal{O}\left( \log^2 \left(\frac{1}{\epsilon}\right) \log N \right)$. The block-encoding of 
\begin{align}
     \frac{1}{ \alpha ||E||_F||D||^p} \sum_{i=1}^N  q(\xbf_i) \ket{i-1} \bra{i-1}
\end{align}
can be obtained via Lemma \ref{lemma: diagonal}, thus incurring a complexity $\mathcal{O}\left(\log^2 \left(\frac{1}{\epsilon}\right) \log N  \right)$. Next, we build the block-encoding of 
\begin{align}
    \sum_{i,j=1}^N \frac{q_{\min}}{ q(\xbf_i) q(\xbf_j)} \ket{i-1}\bra{j-1}
\end{align}
which involves Lemma \ref{lemma: tensorproduct} and Lemma \ref{lemma: negative}, resulting in the complexity:
\begin{align}
    \mathcal{O}\left( \frac{1}{q_{\min}} \log^2 \left(\frac{1}{\epsilon}\right) \log (N) \log^2 \frac{\alpha ||E||_F||D||^p }{\epsilon}  \right)
\end{align}
Next, from the above block-encoded operator, we construct the block-encoding of:
\begin{align}
   \sum_{i,j=1}^N \frac{ w_{\min} }{ w(\xbf_i) }\ket{i-1}\bra{i-1}
\end{align}
which involves the use of Lemma \ref{lemma: product} and Lemma \ref{lemma: negative} again, resulting in the complexity:
\begin{align}
    \mathcal{O}\left( \frac{1}{ w_{\min}q_{\min}} \log^2 \left(\frac{1}{\epsilon}\right) \log (N) \log^2 \left( \frac{\alpha ||E||_F||D||^p }{\epsilon} \right) \log^2 \left( \frac{w_{\min}}{\epsilon} \right) \right)
\end{align}
Next, the construction of the block-encoding of:
\begin{align}
   \left( \frac{w_{\min} q_{\min}}{ \alpha||D||^p}\right)^2 P^\dagger P
\end{align}
requires an application of Lemma \ref{lemma: product} and Lemma \ref{lemma: improveddme}, which leads to the total complexity:
\begin{align}
    \mathcal{O}\left( \frac{1}{ w_{\min}q_{\min}} \log^2 \left(\frac{1}{\epsilon}\right) \log (N) \log^2 \left( \frac{\alpha ||E||_F||D||^p }{\epsilon} \right) \log^2 \left( \frac{w_{\min}}{\epsilon} \right) \right)
\end{align}
The application of Lemma \ref{lemma: positive} to obtain the block-encoding of
\begin{align}
   \left( \frac{w_{\min} q_{\min}}{ \alpha||D||^p}\right) P
\end{align}
incurs a total complexity 
\begin{align}
    \mathcal{O}\left( \frac{1}{ w_{\min}q_{\min}} \log^2 \left(\frac{1}{\epsilon}\right) \log (N) \log^2 \left( \frac{\alpha ||E||_F||D||^p }{\epsilon} \right) \log^2 \left( \frac{w_{\min}}{\epsilon} \right) \log^2 \left( \frac{\lambda_{\min}}{\epsilon} \right) \right)
\end{align}
where $\lambda_{\min} = \min \{ \lambda_1,\lambda_2, ... , \lambda_N  \}$. The final step is to use Lemma \ref{lemma: largesteigenvalues} to find $n$ largest components, with accuracy $\frac{w_{\min} q_{\min}}{ \alpha||D||^p} \epsilon$ as pointed out before. The total complexity is then:
\begin{align}
    \mathcal{O}\left( \frac{\alpha ||D||^p}{ w^2_{\min}q^2_{\min}} \log^2 \left(\frac{1}{\epsilon}\right) \log (N) \log^2 \left( \frac{\alpha ||E||_F||D||^p }{\epsilon} \right) \log^2 \left( \frac{w_{\min}}{\epsilon} \right) \log^2 \left( \frac{\lambda_{\min}}{\epsilon} \right) \log^m \left(\frac{N}{\epsilon}\right) \frac{1}{\Delta^n \epsilon}\log^n \left( \frac{1}{\epsilon}\right) \right)
\end{align}
where $\Delta$ in this case is defined as the $\max \{ |\lambda_{i}-\lambda_{i+1}| \}_{i=1}^n$. The above complexity can be further simplified (asymptotically) as:
\begin{align}
      \mathcal{O}\left( \frac{\alpha ||D||^p}{ w^2_{\min}q^2_{\min}\epsilon} \left( \log^{n+1} (N)+  \log^{2n+6} \left( \frac{1}{\epsilon} \right) \right)\right)
\end{align}
We also recall that:
\begin{align}
     ||D|| &= \sqrt{\sum_{i,j=1}^N d(\xbf_i,\xbf_j)^2}\\
     q(\xbf_i) &= \sum_{j=1}^N K_{ij} = \sum_{j=1}^N \exp\left( - \frac{d(\xbf_i,\xbf_j)^2}{\sigma^2}\right) \\
     q_{\min} &= \min \{ q(\xbf_i)q(\xbf_j) \}_{i,j=1}^N  \\
     w(\xbf_i) &= \sum_{j=1}^N  \frac{K_{ij}}{q(\xbf_i) q(\xbf_j)} \\
     w_{\min} &= \min \{  w(\xbf_i) \} \\
     p &=\mathcal{O}\left( \log \frac{1}{\epsilon}\right) 
\end{align}
where $p$ is the order of the polynomial used to approximate the function $\exp(-x^2)$. Thus, the performance of our algorithm depends quite critically on the net values of the pairwise distances $\{ d(\xbf_i,\xbf_j)\}_{i,j=1}^N$. In the best case, if all $  w_{\min}, q_{\min}, ||D|| =\mathcal{O}(1) $ then our algorithm achieves a polylogarithmic scaling of scaling in the number of data points $N$ and the inverse of error tolerance $\frac{1}{\epsilon}$.

\section{A review of quantum PCA underlying Lemma \ref{lemma: largesteigenvalues}}
\label{sec: reviewPCA}
We provide a more details of Lemma \ref{lemma: largesteigenvalues}, which is essentially the quantum PCA/power method proposed in \cite{nghiem2025refined,nghiem2025improved, nghiem2023improved}. We further note that the power method is also recently employed in \cite{chen2025quantum}. A more thorough discussion can be found in the Appendix F of \cite{nghiem2025improved}. 

First we recall the following recipes in \cite{gilyen2019quantum}:
\begin{lemma}[Corollary 64 of \cite{gilyen2019quantum}  ]
\label{lemma: exponential}
   Let $\beta \in \mathbb{R}_+$ and $\epsilon \in (0,1/2]$. There exists an efficiently constructible polynomial $P \in \mathbb{R}[x]$ such that 
   $$ \Big|\!\Big| e^{ -\beta ( 1-x ) } - P(x)  \Big|\!\Big|_{x\in[-1,1]} \leq \epsilon. $$
   Moreover, the degree of $P$ is $\mathcal{O}\Big( \sqrt{\max[\beta, \log(\frac{1}{\epsilon})] \log(\frac{1}{\epsilon}}) \Big).$
\end{lemma}

\begin{lemma}
\label{lemma: qsvt}[\cite{gilyen2019quantum} Theorem 56]
Suppose that $U$ is an
$(\alpha, a, \epsilon)$-encoding of a Hermitian matrix $A$. (See Definition 43 of~\cite{gilyen2019quantum} for the definition.)
If $P \in \mathbb{R}[x]$ is a degree-$d$ polynomial satisfying that
\begin{itemize}
\item for all $x \in [-1,1]$: $|P(x)| \leq \frac{1}{2}$,
\end{itemize}
then, there is a quantum circuit $\tilde{U}$, which is an $(1,a+2,4d \sqrt{\frac{\epsilon}{\alpha}})$-encoding of $P(A/\alpha)$ and
consists of $d$ applications of $U$ and $U^\dagger$ gates, a single application of controlled-$U$ and $\mathcal{O}((a+1)d)$
other one- and two-qubit gates.
\end{lemma}
Let $\{ \lambda_i, \ket{\lambda_i}\}$ denotes the eigenvalues and corresponding eigenvectors of the given matrix $A$. Let $U_A$ denote the unitary block encoding of $A$. Assume WOLG that the eigenvalues have transcending order $\lambda_1 > \lambda_2 >  ... $. The procedure for finding $k$ largest eigenvalues/eigenvectors is summarized as follows.
\begin{enumerate}
    \item Use Lemma \ref{lemma: product} $k$ times to constrct construct the block encoding of $A^k$. Let $\ket{x_0}$ denote some initial state, generated by some known circuit $U_0$ (assuming to have $\mathcal{O}(1)$ depth). Defined $x_k = A^k \ket{x_0}$ and the normalized state $\ket{x_k} = \frac{x_k}{||x_k||}$. 
    \item Use the block encoding of $A^k$ to apply it to $\ket{x_0}$, we obtain the state:
\begin{align}
   \ket{\phi_1} =  \ket{\bf 0} A^k \ket{x_0} + \ket{\rm Garbage}
\end{align}
   \item Use Lemma \ref{lemma: improveddme} allows us to construct the block encoding of $\ket{\phi_1}\bra{\phi_1}$, which is exactly the block encoding of $x_k x_k^\dagger = ||x_k||^2 \ket{x_k} \bra{x_k} $, according to the \ref{def: blockencode}. 
  \item Define $\gamma \equiv ||x_k||^2 $. We use Lemma \ref{lemma: qsvt} and \ref{lemma: exponential} to transform the block-encoded operator: 
\begin{align}
    \gamma \ket{x_k}\bra{x_k} \longrightarrow e^{-\beta(1-\gamma)} \ket{x_k}\bra{x_k}
\end{align}
\item Recall that we are given $U_0$ that generates the state $\ket{x_0}$, \ref{lemma: improveddme} allows us to block-encode the operator $\ket{x_0}\bra{x_0}$. Now we take the above block encoding and apply it to $\ket{x_0}$, and according to \ref{def: blockencode}, we obtain the following state:
\begin{align}
 \ket{\bf 0} \braket{x_k,x_0} e^{-\beta(1-\gamma)} \ket{x_k} + \ket{\rm Garbage}
    \label{eqn: d4}
\end{align}
\item  Measuring the first register and post-select on $\ket{\bf 0}$, yields the state $\ket{x_k}$ on the remaining register. The success probability of this measurement is $|\braket{x_k,x_0}|^2 e^{-2\beta(1-\gamma)}$, which can be improved quadratically better using amplitude amplification. By choosing $\beta$ sufficiently small, the value of $e^{-2\beta(1-\gamma)} $ is lower bounded by some constant, e.g., $1/2$, thus the probability can be lower bounded by $ \frac{1}{2}|\braket{x_k,x_0}|  $. We note that the overlaps above can be estimated via Hadamard test or SWAP test.
\item From $\ket{x_k}$, we use the block encoding of $A$ to apply and obtain the state:
\begin{align}
    \ket{\bf 0} A \ket{x_k} + \ket{\rm Garbage}
\end{align}
Taking another copy of $\ket{x_k}$ and append another ancilla $\ket{\bf 0}$, we then observe that the overlaps: 
\begin{align}
    \bra{\bf 0}\bra{x_k} \big( \ket{\bf 0} A \ket{x_k} + \ket{\rm Garbage}\big) = \bra{x_k}A \ket{x_k}
\end{align}
which is an approximation to the largest eigenvalue of $A$. According to \cite{friedman1998error, golub2013matrix1}, the value of $k$ needs to be of order $\mathcal{O}\big( \frac{1}{\Delta} (\log \frac{n}{\epsilon} \big)$ to achieve a $\epsilon$ additive accurac, i.e.,
\begin{align}
    |\bra{x_k}A\ket{x_k} - A_1| \leq \epsilon \\
    || \ket{x_k} - \ket{A_1} || \leq \epsilon
\end{align}
\end{enumerate}
The total complexity for estimating largest eigenvalue $\lambda_1$, up to $\epsilon$ error is 
$$\mathcal{O}\Big( \frac{1}{\Delta \Gamma \epsilon} T_A \big(\log \frac{n}{\epsilon}\big) \log\frac{1}{\epsilon}\Big) $$
and the complexity for obtaining $\ket{x_k}$, which is an approximation to $\ket{\lambda_1}$ is $ \mathcal{O}\Big( \frac{1}{\Delta \Gamma} T_A \big(\log \frac{n}{\epsilon}\big) \log \frac{1}{\epsilon} \Big)$, where we have defined $\Gamma \equiv |\braket{x_k,x_0}|$ which is the overlaps between the initially random state and the target state. 

To obtain the operator $\lambda_1 \ket{\lambda_1}\bra{\lambda_1}$, recall from~\ref{eqn: d4} above that we obtained the state: 
\begin{align}
    \ket{\bf 0} \braket{x_k,x_0} e^{-\beta(1-\gamma)} \ket{x_k} + \ket{\rm Garbage} \equiv \ket{\phi}
\end{align}
\ref{lemma: improveddme} allows us to construct the block encoding of $\ket{\phi}\bra{\phi}$, which is:  the block encoding of $|\braket{x_k,x_0}|^2e^{-2\beta(1-\gamma)} \ket{x_k}\bra{x_k} $, and the factor $ |\braket{x_k,x_0}|^2$ can be removed using \ref{lemma: amp_amp}. To proceed, the following result is proved in the Appendix F of \cite{nghiem2025improved}:
\begin{lemma}
    For $\beta \leq \frac{1}{2(1-\gamma)}\log \frac{1}{1-\epsilon}$, we have:
    \begin{align}
        1- e^{-2\beta(1-\gamma)} \leq \epsilon
    \end{align}
\end{lemma}
The above lemma implies:
\begin{align}
    || \ket{x_k}\bra{x_k} - e^{-2\beta(1-\gamma)} \ket{x_k}\bra{x_k} || \leq |1-  e^{-2\beta(1-\gamma)} | \leq \epsilon
\end{align}
So the block-encoded operator $e^{-2\beta(1-\gamma)} \ket{x_k}\bra{x_k}  $ is $\epsilon$-approximated to $\ket{x_k}\bra{x_k}$, which is again an $\epsilon$-approximation of $\ket{\lambda_1}\bra{\lambda_1}$ provided $k$ is chosen properly, as mentioned in the previous paragraph. By additivity, $ e^{-2\beta(1-\gamma)} \ket{x_k}\bra{x_k} $ is $2\epsilon$-approximation to $\ket{\lambda_1}\bra{\lambda_1}$. From the block encoding of $e^{-2\beta(1-\gamma)} \ket{x_k}\bra{x_k} $, we can use \ref{lemma: product} to construct the block encoding of $ A e^{-2\beta(1-\gamma)} \ket{x_k}\bra{x_k} \approx A \ket{\lambda_1}\bra{\lambda_1} = \lambda_1 \ket{\lambda_1}\bra{\lambda_1} $.

To find the second largest eigenvalue/eigenvector $\lambda_2, \ket{\lambda_2}$, we consider the following operator:
\begin{align}
    A - \lambda_1\ket{\lambda_1}\bra{\lambda_1} 
\end{align}
This operator has $\lambda_2,\ket{\lambda_2}$ as the largest eigenvalue/eigenvector, therefore, we can use the above procedure in a straightforward manner. The block-encoding of the above operator can be obtained by using Lemma \ref{lemma: sumencoding} with the block-encoding of $A$ and $\lambda_1 \ket{\lambda_1}\bra{\lambda_2}$. The process is repeated similarly to find the third largest eigenvalue/eigenvector $\lambda_3/\ket{\lambda_3}$.

\section{Block-encoding and quantum singular value transformation}
\label{sec: summaryofnecessarytechniques}
We introduce the main quantum ingredients required for the construction of our algorithm. For brevity, we recapitulate only the key results and omit technical details, which are thoroughly presented in~\cite{gilyen2019quantum}.

\begin{definition}[Block-encoding unitary, see e.g.~\cite{low2017optimal, low2019hamiltonian, gilyen2019quantum}]
\label{def: blockencode} 
Let $A$ be a Hermitian matrix of size $N \times N$ with operator norm $\norm{A} < 1$. A unitary matrix $U$ is said to be an \emph{exact block encoding} of $A$ if
\begin{align}
    U = \begin{pmatrix}
       A & * \\
       * & * \\
    \end{pmatrix},
\end{align}
where the top-left block of $U$ corresponds to $A$. Equivalently, one can write
\begin{equation}
    U = \ket{\mathbf{0}}\bra{\mathbf{0}} \otimes A + (\cdots),    
\end{equation}
where $\ket{\mathbf{0}}$ denotes an ancillary state used for block encoding, and $(\cdots)$ represents the remaining components orthogonal to $\ket{\mathbf{0}}\bra{\mathbf{0}} \otimes A$. If instead $U$ satisfies
\begin{equation}
    U = \ket{\mathbf{0}}\bra{\mathbf{0}} \otimes \tilde{A} + (\cdots),
\end{equation}
for some $\tilde{A}$ such that $\|\tilde{A} - A\| \leq \epsilon$, then $U$ is called an {$\epsilon$-approximate block encoding} of $A$. Furthermore, the action of $U$ on a state $\ket{\mathbf{0}}\ket{\phi}$ is given by
\begin{align}
    \label{eqn: action}
    U \ket{\mathbf{0}}\ket{\phi} = \ket{\mathbf{0}} A\ket{\phi} + \ket{\mathrm{garbage}},
\end{align}
where $\ket{\mathrm{garbage}}$ is a state orthogonal to $\ket{\mathbf{0}}A\ket{\phi}$. The circuit complexity (e.g., depth) of $U$ is referred to as the {complexity of block encoding $A$}.
\end{definition}

Based on Definition~\ref{def: blockencode}, several properties, though immediate, are of particular importance and are listed below.
\begin{remark}[Properties of block-encoding unitary]
The block-encoding framework has the following immediate consequences:
\begin{enumerate}[label=(\roman*)]
    \item Any unitary $U$ is trivially an exact block encoding of itself.
    \item If $U$ is a block encoding of $A$, then so is $\Ibb_m \otimes U$ for any $m \geq 1$.
    \item The identity matrix $\Ibb_m$ can be trivially block encoded, for example, by $\sigma_z \otimes \Ibb_m$.
\end{enumerate}
\end{remark}

Given a set of block-encoded operators, a variety of arithmetic operations can be performed on them. In the following, we present several operations that are particularly relevant and important to our algorithm. Here, we omit the proofs and focus on the implementation aspects, particularly the time complexity. Detailed discussions can be found, for instance, in~\cite{gilyen2019quantum, camps2020approximate}.

\begin{lemma}[Informal, product of block-encoded operators, see e.g.~\cite{gilyen2019quantum}]
\label{lemma: product}
    Given unitary block encodings of two matrices $A_1$ and $A_2$, with respective implementation complexities $T_1$ and $T_2$, there exists an efficient procedure for constructing a unitary block encoding of the product $A_1 A_2$ with complexity $T_1 + T_2$.
\end{lemma}

\begin{lemma}[Informal, tensor product of block-encoded operators, see e.g.~{\cite[Theorem 1]{camps2020approximate}}]\label{lemma: tensorproduct}
    Given unitary block-encodings $\{U_i\}_{i=1}^m$ of multiple operators $\{M_i\}_{i=1}^m$ (assumed to be exact), there exists a procedure that constructs a unitary block-encoding of $\bigotimes_{i=1}^m M_i$ using a single application of each $U_i$ and $\mathcal{O}(1)$ SWAP gates.
\end{lemma}

\begin{lemma}[Informal, linear combination of block-encoded operators, see e.g.~{\cite[Theorem 52]{gilyen2019quantum}}]
    Given the unitary block encoding of multiple operators $\{A_i\}_{i=1}^m$. Then, there is a procedure that produces a unitary block encoding operator of $\sum_{i=1}^m \pm (A_i/m) $ in time complexity $\mathcal{O}(m)$, e.g., using the block encoding of each operator $A_i$ a single time. 
    \label{lemma: sumencoding}
\end{lemma}

\begin{lemma}[Informal, Scaling multiplication of block-encoded operators] 
\label{lemma: scale}
    Given a block encoding of some matrix $A$, as in Definition~\ref{def: blockencode}, the block encoding of $A/p$ where $p > 1$ can be prepared with an extra $\mathcal{O}(1)$ cost.
\end{lemma}

To show this, we note that the matrix representation of the $R_Y$ rotation gate is given by
\begin{equation}
    R_Y(\theta) = \begin{pmatrix}
        \cos(\theta/2) & -\sin(\theta/2) \\
        \sin(\theta/2) & \cos(\theta/2) 
    \end{pmatrix}. 
\end{equation}

If we choose $\theta=2\cos^{-1}(1/p)$, then by Lemma~\ref{lemma: tensorproduct}, we can construct a block-encoding of $R_Y(\theta) \otimes \mathbb{I}_{{\rm dim}(A)}$, where ${\rm dim}(A)$ refers to the dimension of the rows (or columns) of the square matrix $A$. This operation results in a diagonal matrix of size ${\rm dim}(A) \times {\rm dim}(A)$ with all diagonal entries equal to $1/p$. Then, by applying Lemma~\ref{lemma: product}, we can construct a block-encoding of
\begin{equation}
     \frac{1}{p} \ \mathbb{I}_{{\rm dim}(A)} \cdot A = \frac{A}{p}
\end{equation}

\begin{lemma}[Matrix inversion, see e.g.~\cite{gilyen2019quantum, childs2017quantum}]\label{lemma: matrixinversion}
Given a block encoding of some matrix $A$  with operator norm $||A|| \leq 1$ and block-encoding complexity $T_A$, then there is a quantum circuit producing an $\epsilon$-approximated block encoding of ${A^{-1}}/{\kappa}$ where $\kappa$ is the conditional number of $A$. The complexity of this quantum circuit is $\mathcal{O}\left( \kappa T_A \log \left({1}/{\epsilon}\right)\right)$. 
\end{lemma}

\begin{lemma}\label{lemma: amp_amp}[\cite{gilyen2019quantum} Theorem 30]
\label{lemma: amplification}
Let $U$, $\Pi$, $\widetilde{\Pi} \in {\rm End}(\mathcal{H}_U)$ be linear operators on $\mathcal{H}_U$ such that $U$ is a unitary, and $\Pi$, $\widetilde{\Pi}$ are orthogonal projectors. 
Let $\gamma>1$ and $\delta,\epsilon \in (0,\frac{1}{2})$. 
Suppose that $\widetilde{\Pi}U\Pi=W \Sigma V^\dagger=\sum_{i}\varsigma_i\ket{w_i}\bra{v_i}$ is a singular value decomposition. 
Then there is an $m= \mathcal{O} \Big(\frac{\gamma}{\delta}
\log \left(\frac{\gamma}{\epsilon} \right)\Big)$ and an efficiently computable $\Phi\in\mathbb{R}^m$ such that
\begin{equation}
\left(\bra{+}\otimes\widetilde{\Pi}_{\leq\frac{1-\delta}{\gamma}}\right)U_\Phi \left(\ket{+}\otimes\Pi_{\leq\frac{1-\delta}{\gamma}}\right)=\sum_{i\colon\varsigma_i\leq \frac{1-\delta}{\gamma} }\tilde{\varsigma}_i\ket{w_i}\bra{v_i} , \text{ where } \Big|\!\Big|\frac{\tilde{\varsigma}_i}{\gamma\varsigma_i}-1 \Big|\!\Big|\leq \epsilon.
\end{equation}
Moreover, $U_\Phi$ can be implemented using a single ancilla qubit with $m$ uses of $U$ and $U^\dagger$, $m$ uses of C$_\Pi$NOT and $m$ uses of C$_{\widetilde{\Pi}}$NOT gates and $m$ single qubit gates.
Here,
\begin{itemize}
\item C$_\Pi$NOT$:=X \otimes \Pi + I \otimes (I - \Pi)$ and a similar definition for C$_{\widetilde{\Pi}}$NOT; see Definition 2 in \cite{gilyen2019quantum},
\item $U_\Phi$: alternating phase modulation sequence; see Definition 15 in \cite{gilyen2019quantum},
\item $\Pi_{\leq \delta}$, $\widetilde{\Pi}_{\leq \delta}$: singular value threshold projectors; see Definition 24 in \cite{gilyen2019quantum}.
\end{itemize}
\end{lemma}

\section{A Review of Differential Geometry}
\label{sec: reviewofdifferentialgeomtry}
In this section, we provide a brief review of differential geometry. We quote many definitions and related concepts from \cite{nakahara2018geometry}. A more detailed treatment of the subject can be found in the standard literature, e.g., \cite{nakahara2018geometry, lee2009manifolds}.

\begin{definition}[Topological space]
A topological space $X$ is a set $X$ together with a collection $\mathcal{C}$ of subsets of $X$, called \textit{open sets} that satisfies the four axioms:
\begin{enumerate}
    \item $X \in \mathcal{C}$.
    \item The empty set $\varnothing \in \mathcal{C}$.
    \item The intersection of a finite number of elements of $\mathcal{C}$ belongs to $\mathcal{C}$, i.e. $\bigcap_{i=1}^n S_i \in \mathcal{C}$ where $S_i\in \mathcal{C}$ and $n\in\mathbb{Z}_{\ge 1}$.
    \item The union of an arbitrary number of elements of $\mathcal{C}$ belongs to $\mathcal{C}$, i.e. $\bigcup_{i\in T} S_i \in \mathcal{C}$ where $S_i\in \mathcal{C}$ and $T$ is any index set (finite or infinite). 
\end{enumerate}
\end{definition}

\begin{definition}[Smooth manifold]
$M$ is an $m$-dimensional smooth manifold if:
\begin{enumerate}
    \item $M$ is a topological space, Hausdorff, second-countable and paracompact. (Some technical conditions allow the space to behave nicely; a first-time reader can ignore those conditions, which we include here for completeness).
    \item $M$ is provided with a collection \{($U_i, \varphi_i$)\} where $U_i$ form an open cover of $M$, that is $\cup_i U_i$ = M, and $\varphi_i$  is a homeomorphism from $U_i$ onto an open subset $V_i$ of $\mathbb{R}^m$. The pair $(U_i, \varphi_i)$ is called a \textbf{chart}. 
    \item Given that $U_i \cap U_j \neq \varnothing $, the map $\psi_{ij} = \varphi_i \circ \varphi_j^{-1} $ from $\varphi_j (U_i \cap U_j) $ to $\varphi_i (U_i \cap U_j) $  is smooth ($C^\infty$).
\end{enumerate}
\end{definition}
The collection of such charts is called an \textbf{atlas}. The homeomorphism $\varphi_i$ is represented by $m$-functions $\{x^1(p),x^2(p), ..., x^m(p) \}$, referred to as \textbf{local
coordinates}. If the union of two atlases $\{(U_i, \varphi_i)\}$ and $\{(V_j, \psi_j)\}$ is again an atlas, these atlases are said to be \textbf{compatible}. The compatibility is an  equivalence relation, for which the equivalence class is called the \textbf{differentiable structure}. As a concluding remark, differentiable manifold is classified based on differentiable structure. 

\begin{definition}[Tangent space]
In a manifold $M$, a vector is defined to be a \textbf{tangent vector} to a curve in $M$. If a vector is smoothly assigned to each point of $M$, it is called a \textbf{vector field} over $M$. Given $p\in M$, the collection of tangent vectors forms a vector space, called the \textbf{tangent space} at $p$, denoted $T_pM$. It can be shown that dim($T_pM$) = dim(M). The basis of such space is the \textbf{unit-vectors} $\{ \partial/\partial x^\mu \}$, where $\{x^{\mu}\}$ is local coordinate of the point $p$.
\end{definition}

Given $2$ smooth manifolds $M$ and $\tilde{M}$, equipped with atlases $\{(U_i,\varphi_i)\}$ and $\{(\tilde{U}_i,\tilde{\varphi}_i)\}$ respectively, a continuous map $f:M\rightarrow \tilde{M}$ is \textbf{smooth} if $\tilde{\varphi}_j\circ f\circ {\varphi_i}^{-1}$ is smooth for all charts where this mapping is defined.

\begin{definition}[Immersion, submanifold, embedding]
Let $f: M \rightarrow N$ be a smooth map. 
\begin{enumerate}
    \item The map $f$ is called an \textbf{immersion} of $M$ into $N$ if: $f_*: T_pM \rightarrow T_{f(p)}N$ is an injection (one to one), that is: rank($f_*$) = dim $M$. Note that this forces $\dim M\le \dim N$.
    \item The map $f$ is called an \textbf{embedding} if $f$ is an injection and an immersion. The image $f(M)$ is called a \textbf{submanifold} of $N$. Thus, $f(M)$ is diffeomorphic to $M$.
\end{enumerate}
\end{definition}

\begin{definition}[Cotangent space]
    $T_pM$ defines a vector space at each point $p$ of $M$, and hence, there exist a dual space $\mathrm{Hom}(T_pM,\mathbb{R})$, called the \textbf{cotangent space} at $p$, denoted as $T^*_pM$. An element of $T^*_pM$ is a linear map from $T_pM$ to $\Rbb$, called a \textbf{one-form}.
    The dual space $T^*_pM$ forms a linear space, or space of one-form; the basis of such space is the \textbf{unit-covectors} $\{dx^{\mu}\}$.
\end{definition}

\begin{definition}
A tensor of type $(q,r)$ is a multilinear map which maps $q$ elements of $T_pM$ and $r$ elements of $T_p^*M$ to $\mathbb{R}$. The collection of such type $(q,r)$ tensors forms a vector space. An element of the space is written in terms of the bases as: 
\begin{align}
    T = T^{\mu_1 \mu_2...\mu_q}_{\nu_1 \nu_2...\nu_r}(x) \frac{\partial}{\partial x^{\mu_1}} \otimes \frac{\partial}{\partial x^{\mu_2}}  \otimes  ...\otimes \frac{\partial}{\partial x^{\mu_q}} \otimes dx^{\nu_1} \otimes dx^{\nu_2} \otimes ... \otimes dx^{\nu_r} \ ,
\end{align}
where $T^{\mu_1 \mu_2...\mu_q}_{\nu_1 \nu_2...\nu_r}(x)$ is a function (a component of the tensor $T$), $\otimes$ is the \textbf{tensor product} (which keeps the ordering meaningful, so $dx^{\mu} \otimes dx^{\nu} \neq dx^{\nu} \otimes dx^{\mu}$ in general).
\end{definition}
Note that throughout this text, we adopt the Einstein summation convention -- whenever an index (e.g. $\mu)$ appears exactly twice, it is understood to be summed over its full range, so the explicit summing over symbol (e.g. $\sum_\mu$) is omitted.

\begin{definition}[Differential Form]
A \textbf{differential form} of order $r$ or an \textbf{r-form} is a totally anti-symmetric tensor of type $(0,r)$. We can build an $r$-form from the \textbf{wedge product} $\wedge$ of $r$ one-forms, in which the wedge product is defined from the tensor product by the following anti-symmetrization: 
\begin{align}
    dx^{\mu_1} \wedge dx^{\mu_2} \wedge ... \wedge dx^{\mu_r} = \frac1{r!} \sum_{P \in \mathcal{S}_r} \mathrm{sgn}(P)\, dx^{\mu_{P(1)}} \otimes dx^{\mu_{P(2)}} \otimes ... \otimes dx^{\mu_{P(r)}}
\end{align}
where $\mathcal{S}_r$ is the symmetric group (all 
$r!$ permutations of $\{1,2,...,r\}$, treated as a bijective function from $\{1,2,...,r\}$ to itself), $P(j)$ is the image of $j$ under a permutation $P$, and $\mathrm{sgn}(P)=+1$ for even permutations and $-1$ for odd ones. 
\end{definition}

For example, the $r=2$ case gives:
\begin{align}
dx^{\mu} \wedge dx^{\nu} = dx^{\mu} \otimes dx^{\nu} - dx^{\nu} \otimes dx^{\mu} \ ,
\end{align}
The collection of $r$-forms at $p \in M$ forms a \textbf{vector space} $ \Omega^r_p(M)$. An element of $\Omega^r_p(M)$ could be expanded as: 
\begin{align}
      \omega = \frac{1}{r!} \omega_{\mu_1\mu_2...\mu_r}(x) dx^{\mu_1} \wedge dx^{\mu_2} \wedge ... \wedge dx^{\mu_r} \ .
\end{align}

\begin{definition}[Riemannian metric] A \textbf{Riemannian metric} on $M$ is a type $(0,2)$ tensor field $g$ such that: at every point $p \in M$ one equips the tangent space $T_p M$ with an inner product $g_p: T_p M \times T_p M\rightarrow \mathbb{R}$, where it satisfies the following axioms: 
\begin{enumerate}
    \item Symmetry: $g_p(U,V) = g_p(V,U)$, for all tangent vectors $U, V \in T_p M$.
    \item Positive-definiteness: $g_p(U,U) \geq 0$, for all $U\in T_p M$, where the equality holds only when $U = 0$.
\end{enumerate}
\end{definition}

In the local coordinate $\{ x^\mu \}$, we can write the inner product between vectors $U$ and $V$ as:
\begin{equation}
    U = U^\mu(x) \frac{\partial}{\partial x^\mu} \ , \ V = V^\nu(x) \frac{\partial}{\partial x^\nu} \ \ \Longrightarrow \ \ g_p(U,V) = g_{\mu\nu}(x) U^\mu(x) V^\nu(x) \ .
\end{equation}
This makes $g_p$ a symmetric (i.e. $g_{\mu\nu}(x)=g_{\nu\mu}(x)$), positive-definite bilinear form on $T_p M$, which also gives rise to an isomorphism between $T_p M $ and $T^*_p M$. To see that, let us define the inverse metric $g^{\mu\nu}(x)$, which is the matrix inverse of $g_{\mu\nu}(x)$:
\begin{equation}
g^{\mu\nu} (x) = g^{-1}_{\mu\nu}(x) \ , \ g^{\mu\nu}(x) g_{\nu\rho}(x) = \delta^{\mu}_\rho \ ,
\end{equation}
where \(\delta^{\mu}_{\rho}\) is the Kronecker delta. Because $g_{\mu\nu}(x)$ is symmetric and positive-definite at every point, its inverse exists and is also symmetric, and hence $g^{\mu\nu}(x) = g^{\nu\mu}(x)$. We can then define the \textbf{musical isomorphisms}:
\begin{itemize}
    \item Raising-index (sharp $\sharp$) map with the inverse metric tensor $g^{\mu\nu}(x)$: 
    \begin{align}
    ( \cdot )^\sharp \ : \ T^*_p M \rightarrow T_p M \ \ \text{, e.g.} \ \ U_\mu(x) \rightarrow U^{\sharp \mu}(x) = g^{\mu \nu} (x) U_\nu (x) \ .
    \end{align}
    \item Lowering-index (flat $\flat$) map with the inverse metric tensor $g^{\mu\nu}(x)$:
    \begin{align}
    ( \cdot )_\flat \ : \ T_p M \rightarrow T^*_p M \ \ \text{, e.g.} \ \ U^\mu(x) \rightarrow U_{\flat \mu}(x) = g_{\mu \nu} (x) U^\nu (x) \ .
    \end{align}
\end{itemize}
From these definitions, we get $(U^\sharp)_\flat = (U_\flat)^\sharp = U$, which provides a canonical correspondence between vectors and covectors induced by the metric. 

The same index–raising and index–lowering operations extend component-wise to
tensors of any type: given a $(q,r)$ tensor $T$, we can raise or lower an index by contracting it with the metric or its
inverse, allowing a metric‐induced isomorphism between the spaces of all $(q,r)$ and $(q\pm 1, r\mp 1)$ tensors.

\begin{definition}[Line-element] Because the metric supplies an inner product on every tangent space, it assigns a squared length to any infinitesimal displacement. If a curve on the manifold $M$ at point $p$ has coordinate differential $\{d x^\mu\}$, its corresponding tangent vector is $U=dx^\mu \partial/\partial x^\mu \in T_p M$ (one should not confuse $U$ with the $(1,1)$ tensor $dx^\mu \otimes \partial/\partial x_\nu$ or its trace $dx^\mu \otimes \partial/\partial x^\mu$). The squared-norm of $U$ under the metric $g$ (which is its own inner product) is given by:
\begin{equation}
    g_p(U,U) = g_{\mu\nu}(x) dx^\mu dx^\nu \ .
\end{equation}
The following expression
\begin{equation}
    ds^2 = g_{\mu\nu}(x) dx^\mu dx^\nu
\end{equation}
is called the \textbf{line element}, which defines a coordinate-free infinitesimal squared distance on the manifold.
\end{definition}

Similarly, we can also define other coordinate-free geometric measurements. For example, the \textbf{volume element}:
\begin{equation}
    dV = \sqrt{\text{det}[g(x)]} dx^1 dx^2 ... dx^m \ ,
\end{equation}
where $m = \text{dim}(M)$. This expression is associated with the following $m$-form (something can be integrated over all $m$ coordinates to obtain a scalar-value):
\begin{equation}
    W = \sqrt{\text{det}[g(x)]} dx^1 \wedge dx^2 \wedge ... \wedge dx^m \ ,
\end{equation}
stays invariant under any coordinate change transformation. This is the canonical volume form associated with the metric.

\begin{definition} [Curvature tensor] Curvature are tensor objects that encapsulate how a Riemannian metric ``bend'' space: they quantify the failure of \textit{parallel transport} to return a vector unchanged after tracing an infinitesimal loop. Although their geometric motivation is rich, for our purposes, here we only present the explicit formulae for a few types of \textbf{curvature tensors} that are relevant to our work. The first one is:
\begin{itemize}
    \item \textbf{Riemann curvature} tensor $R$ (type $(1,3)$):
    \begin{equation}
        R^{\rho}_{\sigma \mu\nu}(x) = \partial_\mu \Gamma^{\rho}_{\nu\sigma}(x) - \partial_\nu \Gamma^{\rho}_{\mu\sigma}(x) + \Gamma^{\rho}_{\mu\lambda}(x) \Gamma^{\lambda}_{\nu\sigma}(x) - \Gamma^{\rho}_{\nu\lambda}(x) \Gamma^{\lambda}_{\mu\sigma}(x) \ ,
    \end{equation}
    in which $\Gamma^{\rho}_{\mu\nu}(x)$ is called the \textbf{Christoffel symbol}:
    \begin{equation}      \Gamma^{\rho}_{\mu\nu}(x) = \frac12 g^{\rho\sigma}(x) \left[ \partial_\mu g_{\sigma \nu}(x) + \partial_\nu g_{\sigma \rho}(x) - \partial_\sigma g_{\mu\nu}(x) \right] \ .
    \end{equation}
     It is often more convenient to work with $R_{\rho\sigma\mu\nu} = g_{\rho \gamma}R^{\gamma}_{\sigma\mu\nu}$ (also called the Riemann curvature) rather than $R^{\rho}_{\sigma \mu\nu}$, which is of type $(0,4)$.
    \end{itemize}
    We shall consider algebraically more simple objects that can be obtained from the Riemann curvature:
    \begin{itemize}
    \item \textbf{Ricci curvature} tensor $C$ (type $(0,2)$):
    \begin{equation}
        C_{\mu\nu}(x) = R^{\rho}_{ \mu \rho\nu}(x) \,
    \end{equation}
    which is a trace of the Riemann tensor. It turns out that the volume element in a curved manifold deviates from that of flat Euclidean space, and this deviation is related to the Ricci curvature, see more from the calculation of Appendix \ref{sec:ball_formula}.
    \item \textbf{Scalar curvature} $S$ (a smooth function from $M$ to $\mathbb{R}$):
    \begin{equation}
        S(x) = g^{\mu\nu}R_{\mu\nu}(x) \,
    \end{equation} which is a trace of the Ricci tensor and gives a scalar measure of curvature at a point.
\end{itemize}
\end{definition} 

Here, we hide under the rug the definition of the \textbf{Levi-Civita connection}, which is a way to take derivatives of vector fields on a (curved) manifold, in a manner that is compatible with the metric (i.e. it preserves lengths and angles as measured by the metric) and has no torsion (i.e., it is symmetric).
The Christoffel symbols are the coordinate components of the Levi-Civita connection.

\begin{definition} [Induced metric]
Consider the manifold $M$ embedded in a higher-dimensional $\mathbb{R}^n$ Euclidean space equipped with the Cartesian coordinates $\{X^a\}$ and metric $\delta_{ab}$ (which is a Kronecker delta matrix). We define a metric $g$ on $M$ so that the length of any curve lying in $M$ equals its length when computed in $\mathbb{R}^n$ (like how distances on a globe differ from flat maps). Expressed in local coordinates $\{x^\mu \}$ on $M$, this metric is called the \textbf{induced metric} and can be calculated with:
\begin{equation}
    g_{\mu\nu}(x) = \delta_{ab} \frac{\partial X^a(x)}{\partial x^\mu} \frac{\partial X^b(x)}{\partial x^\nu} \ ,
\end{equation}
where $\{X^a(x)\}$ are the coordinate functions of the embedding $M \rightarrow \mathbb{R}^n$. One can think of the induced metric as the `pull-back' of the Euclidean metric of $\mathbb{R}^n$.
\end{definition}

For our work, the high-dimensional data is a point-cloud in $\mathbb{R}^n$. Under the manifold hypothesis, we approximate this cloud by a smooth and differentiable manifold $M$ endowed with the induced metric $g$. Our goal is then to study the geometric properties of the resulting Riemannian manifold $(M,g)$ -- such as estimating the intrinsic dimensionality $\text{dim}(M)$ and the local scalar curvature $S$.

\end{document}